\documentclass[sigconf]{acmart}

\AtBeginDocument{%
  \providecommand\BibTeX{{%
    \normalfont B\kern-0.5em{\scshape i\kern-0.25em b}\kern-0.8em\TeX}}}

\copyrightyear{2024}
\acmYear{2024}
\setcopyright{rightsretained}
\acmConference[ICSE '24]{2024 IEEE/ACM 46th International Conference on
Software Engineering}{April 14--20, 2024}{Lisbon, Portugal}
\acmBooktitle{2024 IEEE/ACM 46th International Conference on Software
Engineering (ICSE '24), April 14--20, 2024, Lisbon,
Portugal}\acmDOI{10.1145/3597503.3623345}
\acmISBN{979-8-4007-0217-4/24/04}

\usepackage{amsmath,amsfonts,bm}

\def\eqref#1{equation~\ref{#1}}

\def\1{\bm{1}}

\DeclareMathAlphabet{\mathsfit}{\encodingdefault}{\sfdefault}{m}{sl}
\SetMathAlphabet{\mathsfit}{bold}{\encodingdefault}{\sfdefault}{bx}{n}

\usepackage{hyperref}
\usepackage{url}

\usepackage{amsmath}
\usepackage{graphicx}
\usepackage{tablefootnote}

\usepackage{wrapfig}
\usepackage{svg}
\usepackage{caption}
\usepackage{subcaption}
\usepackage{comment}
\usepackage{makecell}
\usepackage{algorithm}
\usepackage{algpseudocode}
\usepackage{verbatim}
\usepackage{pdfpages}
\usepackage{fancyhdr}
\usepackage{multicol}
\usepackage{color}
\usepackage{siunitx}

\usepackage{appendix}

\usepackage{tcolorbox}

\usepackage{dblfloatfix}
\usepackage{multirow}

\usepackage{booktabs}

\usepackage{totpages}

\begin{document}

\newcommand{\mycomment}[1]{}

\newenvironment{removeenv}{\par\color{gray}}{\par}
\excludecomment{removeenv} 
\newenvironment{wipenv}{\par}{\par}
\newcommand{\wip}[1]{#1}
\newcommand{\todo}[1]{\wip{TODO: #1}}
\newcommand{\needcite}{\wip{(citation needed)}}
\newcommand{\xxx}{\wip{XXX}}
\newcommand{\needcheck}[1]{\textcolor{red}{#1}}

\newcommand{\tool}{DeepDFA}
\newcommand{\dataset}{Big-Vul}
\newcommand{\wei}[1]{{\color{blue}Wei:~[#1]}}
\newcommand{\bent}[1]{{\color{orange}#1}}
\definecolor{cardinal}{rgb}{0.77, 0.12, 0.23}
\definecolor{capri}{rgb}{0.0, 0.75, 1.0}
\newcommand{\ben}[1]{\textcolor{capri}{Ben:~[#1]}}
\newcommand{\finalnew}[1]{#1}
\definecolor{myorange}{RGB}{255, 127, 80}
\newcommand{\hey}[1]{\textcolor{myorange}{#1}}
\newcommand{\ph}{\textcolor{red}{WIP}}
\newcommand{\cmt}[1]{\textcolor{red}{#1}}
\definecolor{asparagus}{rgb}{0.53, 0.66, 0.42}

\newcommand{\citeme}{{{\color{red}(needs citation)}}}
\newcommand{\note}[1]{{{\color{red}Note: #1}}}

\newcommand{\mytablespace}{\cmidrule{2-4}}

\newcommand{\fix}{\marginpar{FIX}}
\newcommand{\new}{\marginpar{NEW}}

\title{Dataflow Analysis-Inspired Deep Learning for \\Efficient Vulnerability Detection}
\renewcommand{\appendixpagename}{Supplemental Materials: Dataflow Analysis-Inspired Deep Learning for Efficient Vulnerability Detection} 



\author{Benjamin Steenhoek}
\affiliation{%
  \institution{Iowa State University}
  \city{Ames}
  \state{Iowa}
  \country{USA}
}
\email{benjis@iastate.edu}

\author{Hongyang Gao}
\affiliation{%
  \institution{Iowa State University}
  \city{Ames}
  \state{Iowa}
  \country{USA}
}
\email{hygao@iastate.edu}

\author{Wei Le}
\affiliation{%
  \institution{Iowa State University}
  \city{Ames}
  \state{Iowa}
  \country{USA}
}
\email{weile@iastate.edu}

%

\begin{abstract}
Deep learning-based vulnerability detection has shown great performance and, in some studies, outperformed static analysis tools. However, the highest-performing approaches use token-based transformer models, which are not the most efficient to capture code semantics required for vulnerability detection. Classical program analysis techniques such as dataflow analysis can detect many types of bugs based on their root causes. In this paper, we propose to combine such causal-based vulnerability detection algorithms with deep learning, aiming to achieve more efficient and effective vulnerability detection. Specifically, we designed {\tool}, a dataflow analysis-inspired graph learning framework and an embedding technique that enables graph learning to simulate dataflow computation.
We show that {\tool} is both performant and efficient.
\tool{} outperformed all non-transformer baselines.
It was trained in 9 minutes, 75x faster than the highest-performing baseline model.
When using only 50+ vulnerable and several hundreds of total examples as training data, the model retained the same performance as 100\% of the dataset. \tool{} also generalized to real-world vulnerabilities in \textsc{DbgBench}; it detected 8.7 out of 17 vulnerabilities on average across folds and was able to distinguish between patched and buggy versions, while the highest-performing baseline models did not detect any vulnerabilities.
By combining \tool{} with a large language model, we surpassed the state-of-the-art vulnerability detection performance on the Big-Vul dataset with 96.46 F1 score, 97.82 precision, and 95.14 recall.
Our replication package is located at \wip{\url{https://doi.org/10.6084/m9.figshare.21225413}}.




\end{abstract}

\maketitle

\title{Dataflow Analysis-Inspired Deep Learning for Efficient Vulnerability Detection} 

\begin{figure*}[ht]
    \centering
    \includegraphics[width=\textwidth]{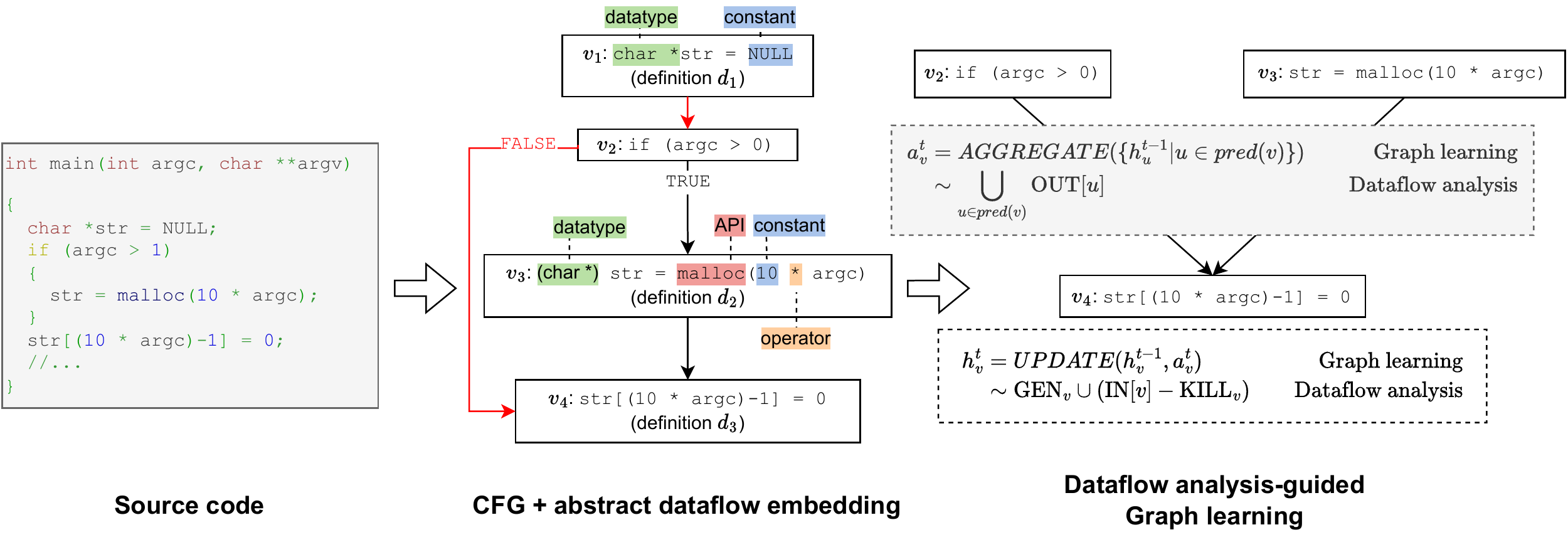}
    \caption{Overview of {\tool}}
    \label{fig:overview}
\end{figure*}

\section{Introduction}
Software vulnerabilities cause great harm to people and corporations.
Many Internet users have had their personal information breached because of security vulnerabilities, with common reports of breaches exposing millions of records~\citep{wikipedia_databreaches_2021}. The average data breach costs the target company \$4.24 million, according to IBM’s 2021 report~\citep{ibm_cost_2021}.
The number of vulnerabilities is growing every year, as reported by the Common Vulnerability Enumeration (CVE) from 2016-2021~\citep{cvedetails_2021}.  Due to its importance, we urgently need to develop effective and automatic vulnerability detection tools.

\definecolor{babyblue}{rgb}{0.54, 0.81, 0.94}

The rapid advance of AI technologies has motivated software companies to invest heavily in deep learning-based vulnerability detection tools~\citep{lu_codexglue_2021,zheng_d2a_2021}.
These tools have outperformed traditional static analysis~\citep{li_vuldeepecker_2018,ding_velvet_2022,cao_mvd_2022}.
Recently, large language models (LLMs) have reported state-of-the-art results; LineVul~\citep{fu_linevul_2022}, a recent model based on CodeBERT, reported 91 F1 score on a commonly used real-world vulnerability dataset~\citep{fan_cc_2020}.

However, LLMs require large amounts of training data and computational resources for training and inference (see \S~\ref{sec:efficiency}), but
a large volume of high-quality vulnerability detection data is hard to get. They also can fail to detect vulnerabilities beyond the training dataset (see \S~\ref{sec:generalization}); for example, the top-performing transformer models LineVul and UniXcoder were not able to detect any of the real-world vulnerabilities in \textsc{DbgBench}~\cite{dbgbench}. 
%
Furthermore, by using solely text tokens, these models may not effectively learn program semantics, such as program values along paths, propagation of taint values, and security-sensitive API calls along the control flow paths. The performance of these models can be further improved when we consider such information (see \S \ref{sec:effectiveness}).


In this paper, we explore the idea of combining {\it dataflow  analysis (DFA)} algorithms with deep learning to develop small, efficient, yet effective models for vulnerability detection.
In prior literature~\cite{xu_how_2019, cranmer_discovering_2020}, deep learning integrated with domain-specific knowledge and algorithms has reported improved performance and better generalization to unseen data, while using less data and computational resources. 

 {\it Dataflow Analysis (DFA)} computes the data usage patterns and relations in the control flow graph (CFG) of a program and reports a vulnerability based on its root cause, i.e., whether the values and data relations collected from the program indicate the occurrence of the vulnerable conditions. {\it Graph learning} ({\it learning based on graph neural networks (GNN)}) can aggregate and propagate information in the graph in a similar fashion to DFA.
 In this paper, we explore the analogy between DFA and the GNN message-passing mechanism and design an embedding technique that encodes dataflow information at each node of the CFG.
 Specifically, we leverage the efficient {\it bit-vector} representation of dataflow facts to encode the definitions and uses of the variables. Graph learning on such an embedding propagates and aggregates dataflow information and thus simulates the dataflow computation as done in DFA. Using this approach, we hope that the learned graph representation can better encode program semantic information, e.g., {\it reaching definitions}, which will be very useful for accurate vulnerability detection.
 
Based on this rationale, we developed an {\it abstract dataflow embedding} that can map variable definitions of individual programs to a common space so that the model can compare and generalize
data usage patterns (dataflow)
related to vulnerabilities across programs. We selected a graph learning architecture whose aggregate and update functions worked most effectively for the dataflow propagation.



Our evaluation shows that DeepDFA is substantially faster than our baseline models in terms of both training and inference time.
It only took 9 minutes to train, and inference on a CPU took 5.8 ms/example. This remarkable efficiency permits applications for personalized training and inference in non-GPU environments.
It is also efficient in its use of training data, achieving its best F1 score using only 50+ vulnerable examples and several hundred total examples (\S~\ref{sec:efficiency}). This frugality allows applications within a single development team, where it may be impractical to collect thousands of vulnerable examples.
Yet, DeepDFA still outperformed all non-transformer baselines (\S~\ref{sec:effectiveness}) and retained its performance on unseen projects better than all baseline models (\S~\ref{sec:generalization}).
Additionally, when applied to a real-world benchmark of unseen projects, {\sc DbgBench}~\cite{dbgbench}, DeepDFA detected 8.7 out of 17 of bugs (averaged over 3 runs) and correctly reported 3 out of 5 patched programs as non-vulnerable (\S~\ref{sec:generalization}). In comparison, the highest-performing baselines, LineVul~\cite{fu_linevul_2022} and UniXcoder~\cite{guo_unixcoder_2022}, did not detect any vulnerabilities.
We also show that DeepDFA's learned representation can be used with other models to further improve their performance.
By combining UniXcoder with DeepDFA, we surpassed state-of-the-art performance with 96.46 F1 score, 97.82 precision, and 95.14 recall.




In summary, we made the following contributions in this paper: 
\begin{enumerate}
    \item We designed an abstract dataflow embedding \finalnew{to enable deep learning to generalize semantics/dataflow patterns of vulnerabilities across programs} (\S \ref{sec:abstract-embedding});
    \item We applied graph learning on the control flow graph (CFG) of the program and abstract dataflow embedding to simulate reaching definition dataflow analysis (\S \ref{sec:dataflow-walkthrough});
    \item We implemented DeepDFA and experimentally demonstrated that {\tool} outperforms baselines in vulnerability detection for effectiveness, efficiency, and generalization over unseen projects (\S \ref{sec:experiments});
    \item We provided rationale to help understand why DeepDFA performs well and is efficient (\S \ref{sec:rationale123}); and
    \item We surpassed the state-of-the-art vulnerability detection performance by combining DeepDFA and UniXcoder (\S \ref{sec:experiments}).
    
    
\end{enumerate}


\begin{figure*}[!ht]
    \newcommand{\myfigureheight}{3cm}
    \centering
    \begin{subfigure}[t]{0.45\textwidth}
        \centering
        \includegraphics[height=\myfigureheight]{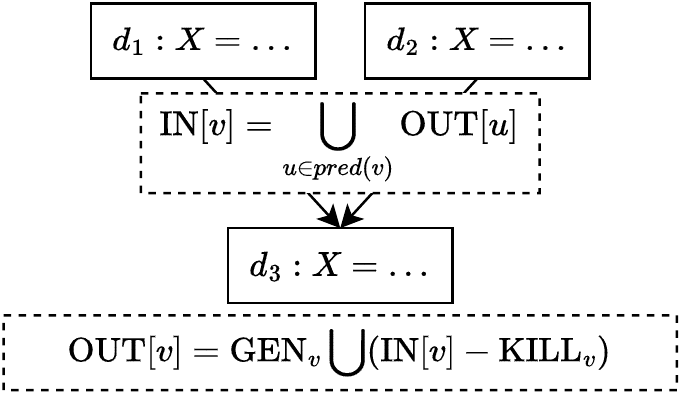}
        \caption{Dataflow Analysis}
        \label{fig:analogy-dataflow}
    \end{subfigure}%
    \begin{subfigure}[t]{0.465\textwidth}
        \centering
        \includegraphics[height=\myfigureheight]{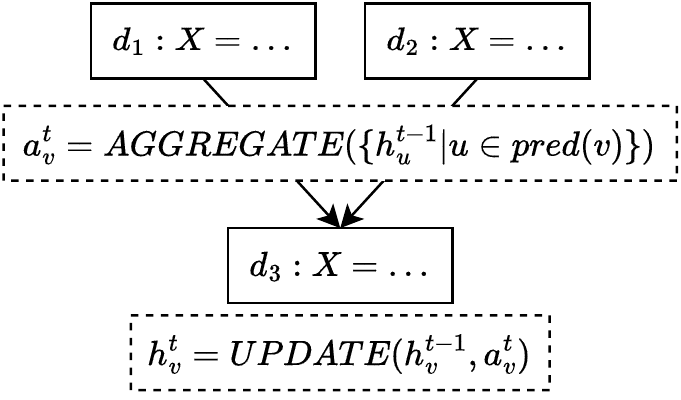}
        \caption{Graph Learning}
        \label{fig:analogy-gnn}
    \end{subfigure}
    \caption{Analogy of information propagation in Dataflow Analysis and Graph Learning}
    \label{fig:analogy}
\end{figure*}

\section{Overview}


\begin{wipenv}
We propose {\tool}, a deep learning framework guided by dataflow analysis algorithms, shown in Figure \ref{fig:overview}.
Given the source code of a potentially vulnerable program (left), we convert it to a CFG and encode the nodes using an \textit{abstract dataflow embedding} which we designed.
The CFG specifies the execution order of statements, and is the data structure on which dataflow analysis operates.

In the middle of the figure, we show our approach of computing abstract dataflow embeddings. In dataflow analysis, definitions of variables, e.g., {\tt a=3}, are program specific. Applying to deep learning, we {\it abstract} these concrete definitions from different programs, and hypothesize that the usage patterns of the {\it abstract definitions} can be compared and summarized across programs during learning. To construct the abstract definitions, we used the properties of definitions that are important for vulnerability detection, based on domain knowledge from program analysis. Specifically, we considered the data types of the defined variable, the API calls, constants, and operators used to define the variables. Inspired by the bit-vector representation used in dataflow analysis, we encode the abstract definitions in a compact and very efficient fashion. 
We will provide more detailed design of this embedding in Section \ref{sec:abstract-embedding}.

We used a bit-vector style of representing a {\it set} of abstraction definitions. This numerical representation can be directly used as the initial node representations for graph learning. In the right of the figure, we apply graph learning which {\it aggregates} the information from nodes like the ``merge'' operation performed in dataflow analysis, and also {\it updates} using the information at each nodes like the ``update'' operation performed in dataflow analysis.
We provide more background on the analogy in Section 3.

Finally, we use the learned graph representation to classify whether the function is vulnerable or not.
By directly propagating dataflow information through graph learning, we hope to present to the classifier a representation of the program which encodes useful information directly related to vulnerability, achieving efficient and effective vulnerability detection. The advantage of deep learning is that the mapping from the encodings of programs to the decisions are learned from the data, but in dataflow analysis, we need to manually craft rules to map from the dataflow analysis results to vulnerability decisions.

\end{wipenv}


\newcommand{\transfer}{f}
\newcommand{\meet}{\sqcap}

\newcommand{\aggregate}{\textit{AGGREGATE}}
\newcommand{\update}{\textit{UPDATE}}

\section{Rationale}\label{sec:rationale123}

In this section, we provide the relevant background of dataflow analysis for vulnerability detection and graph learning. It provides understanding on why our approach is efficient and effective. Then, we compared the closely related work that also considers dataflow in deep learning to clarify the novelty of our work.

\subsection{Dataflow Analysis for Vulnerability Detection}
Dataflow analysis (DFA) is a method for computing data usage patterns in a program. In addition to compiler optimization, dataflow analysis is an important method for vulnerability detection.
One instance of dataflow analysis, called {\it reaching definition analysis}, reports at which program points a particular variable definition can {\it reach}.
A definition {\it reaches} a node when there is a path in the CFG that connects the definition and the node, and the variable is not redefined along the path. The reaching definition analysis can detect a null-pointer dereference vulnerability based on its root cause when it identifies that a definition of an NULL pointer reaches a dereference of the pointer. Similarly, it is a causal step to detect many other vulnerabilities such as buffer overflows, integer overflow,  uninitialized variables, double-free and use-after-free~\citep{cesare_bugalyzecom_2013}. 

DFA uses two equations to propagate the dataflow information through the neighboring nodes in the CFG, namely {\it meet operator} and {\it transfer function}~\citep{aho_compilers_2007}. The meet operator aggregates the dataflow sets from its neighbors.
The transfer function updates the dataflow set using the information available in the node $v$. In the reaching definition analysis, the dataflow set is a set of definitions that reach a program point. A simple approach of performing a DFA is the {\it Kildall method}~\citep{kildall_unified_1973}. It iteratively propagates the dataflow information to the neighbors of $v$ in the CFG, one step at a time.  The algorithm terminates when the dataflow information of all nodes stops changing, denoted a \textit{fixpoint}. At termination, all nodes will incorporate the dataflow information from all other relevant nodes. When used for vulnerability detection, this information is compared to a user-specified vulnerability condition to determine  whether a vulnerability has occurred in the program.

\subsection{Analogy of Graph Learning and Dataflow Analysis}
\label{sec:analogy}

\begin{figure*}[t]
    \centering
    \includegraphics[width=\textwidth]{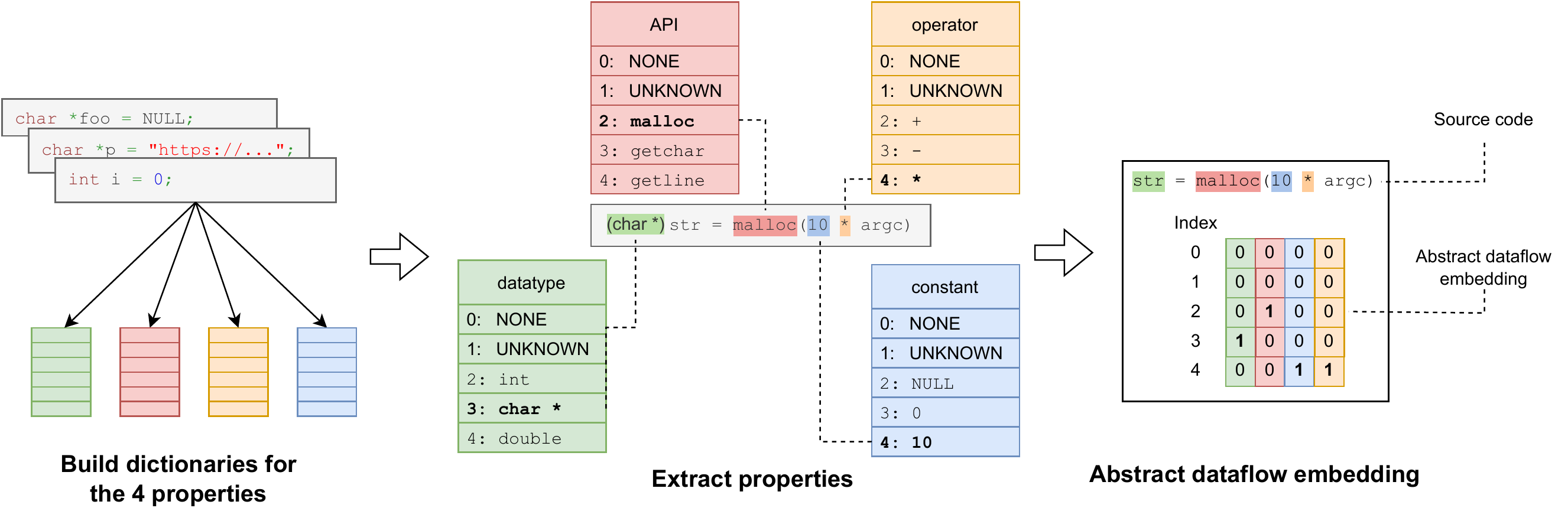}
    \caption{Abstract dataflow embedding generation}
    \label{fig:embedding}
\end{figure*}

Graph learning starts with an initial node representation, and then it performs a fixed number of iterations of the message-passing algorithm~\cite{gilmerNeuralMessagePassing2017a} to propagate information through the graph. The initial node representation is generally a fixed-size continuous vector which represents the content of the node.
At each iteration, each node aggregates information from its neighbors, and then updates its state to integrate the information. The two steps are done through the {\aggregate} and {\update} functions, similar to the two dataflow equations of meet operator and transfer function.
These functions can be simple numerical equations or neural networks. After iteration is done, all node representations are combined to produce a graph-level representation, which is passed to a classifier layer to make a prediction.

In Figure~\ref{fig:analogy}, we visualize the analogy between graph learning and dataflow analysis on a snippet of CFG.
In the CFG, each node is a statement, and each edge indicates the order of execution between two statements.
In Figure \ref{fig:analogy-dataflow}, we show the two dataflow equations~\cite{aho_compilers_2007} that define a reaching definition dataflow analysis. 

\begin{eqnarray}
\label{eq:meet-operator}\text{meet operator:}    & IN[v] = \bigcup_{u \in pred(v)} OUT[u]\\
\label{eq:transfer-function}\text{transfer function:}      & OUT[v] = GEN_v \bigcup  (IN[v]- KILL_v)
\end{eqnarray}

where $IN[v]$ and $OUT[v]$ are the sets of dataflow located at the beginning and end of a statement. $GEN_v$ and $KILL_v$ represent the dataflow {\it generated} (new definitions) and {\it killed} (overwritten definition) in node $v$.
Reaching definition is a {\it may} dataflow problem and thus the meet operator used {\it union} to merge the dataflow information from its predecessors. Meanwhile, reaching definition is a {\it forward} dataflow problem, and thus we used $IN[v]$, $GEN_v$, and $KILL_v$ to compute the dataflow at the exit of the statement.
%

In Figure~\ref{fig:analogy-gnn}, we show an analogous behavior of graph learning.

\begin{eqnarray}
\label{eq:aggregate}\text{aggregate:}  & a^t_v = {\aggregate} (\{ h^{t-1}_u | u \in pred(v) \})\\
\label{eq:update}\text{update:}     & h^t_v = {\update} (h^{t-1}_v, a^t_v)
\end{eqnarray}
where $a^t_v$ denotes the aggregated information from the neighboring nodes and $h^t_v$ denotes the state of node $v$ after $t$ iterations of message-passing (analogous to $OUT[v]$). 
We set $t$ as a hyperparameter.

\subsection{The Novelty of Our Work}\label{sec:novelty}
Previously, researchers have proposed to integrate dataflow information with deep learning for program analysis tasks.
A category of approaches similar to Devign~\cite{zhou_devign_2019} used data dependency graphs as a part of the program representation on which deep learning is performed. However, Devign used word embeddings to encode statements into vector representations based on their unstructured text content. Such an encoding, even propagated through data dependency edges, cannot directly capture the dataflow patterns.

\textsc{ProGraML}~\cite{cummins_programl_2021} has developed graph learning on LLVM IR code and applied it for compiler optimization tasks. It is another work that pointed out the analogy between DFA and graph learning.
Their solution is to modify CFGs by creating instruction nodes and data nodes separately. \textsc{ProGraML} adds control-flow edges between instruction nodes and data-flow edges between the data nodes. However, this work encoded nodes using an embedding which only represents LLVM IR operators and variable types. This approach is very coarse-grained in that many statements can have the same operators and variable types, but they will lead to different dataflow.
Therefore, similar to Devign, the propagation of such an encoding even along dataflow edges does not directly capture dataflow patterns.

Our abstract dataflow embedding attempts to directly represent
the variable definitions which are propagated in DFA and is modeled after the bit-vector representation used in DFA, which allows the network to learn the operations of the dataflow analysis algorithm.
We also target a specific problem (reaching definitions, which was
not targeted by \textsc{ProGraML}), for which the results of DFA are directly useful and pertinent to vulnerability detection (e.g. § 4.2).

\newcommand{\irow}[1]{
  \begin{smallmatrix}[#1]\end{smallmatrix}%
}

\newcommand{\cirow}[1]{\textcolor{red}{\irow{#1}}}

\section{Approach}
\label{sec:approach}
    
Based on the analogous behaviors of DFA and GNN, we designed a node embedding that can represent the dataflow set at each node.
We developed {\tool}, a deep learning framework which
conducts graph learning on the CFG of a program and propagates dataflow information for vulnerability detection.

\subsection{Abstract Dataflow Embedding}
\label{sec:abstract-embedding}

In dataflow analysis, we use a {\it bit vector} to represent the dataflow set at each node.
A bit vector consists of $n$ bits of 0s and 1s. Its length is the size of the domain. A bit is set to 1 if its corresponding element is present in the set. In reaching definition analysis, the domain consists of all the definitions in the program, and the bits are set to ``1'' if the corresponding definitions reach the node. For example, in Figure \ref{fig:overview}, the program contains three definitions at nodes $v_1$, $v_3$, and $v_4$ so the reaching definition analysis uses a bit vector $\irow{0&0&0}$ to initialize each node at the beginning of the analysis. This bit vector represents $OUT[v]$ in the dataflow equations (See Section \ref{sec:analogy}). It is updated at each step of propagation, and when the analysis terminates, the bit vectors for each node represent all possible definitions that can reach {\it that} node.

The bit-vector representation of reaching definition analysis efficiently encodes program semantic features related to vulnerability detection. The definitions of programs can be quickly obtained at the node via lightweight analysis locally at the statements. However, in graph learning, we cannot directly use the bit vector of definitions as the node embedding. This is because in dataflow analysis and the domain of definitions are both specific to a program. In other words, different programs have different variable definitions; the bit vectors of each program thus have different lengths and the elements (each definition) are not comparable either. Whereas, in graph learning, we want to extract dataflow patterns of vulnerabilities from all the programs in the training dataset. Thus, we need to have a ``global'' definition set that can be used to specify definitions for different programs, so that graph learning can compare them and generalize from them.


To address this challenge, we map all the concrete definitions in the programs in a training dataset to {\it abstract definitions} by identifying important properties of the definitions.
Following a list of attack surfaces identified by Moshtari et al.~\cite{moshtari_grounded_2022}, we designed the following four properties that can encompass the attack surfaces of a vulnerability and used them to represent a definition:

\begin{enumerate}
    \item API call: the call to library or system functions used to define a variable, e.g. \texttt{malloc} and \texttt{strlen}.
    \item Data type: the data type of the variable being assigned, e.g. \texttt{int}, \texttt{char*} and {\tt float}.
    \item Constant: the constant values assigned in the definition, e.g. \texttt{NULL}, \texttt{-1} and the hard-coded string \texttt{"foo"}.
    \item Operator: the operators used to define a variable, e.g. \texttt{+}, \texttt{-} and \texttt{*}.
\end{enumerate}

We analyze a large corpus of programs, e.g., the training set, and collect the top-$k$ frequently used API calls, data types, constants and operators to construct a dictionary. $k$ is a hyperparameter of DeepDFA.
We select only the top-$k$ keys because the representations of user-defined names of APIs and data types cannot be generalized across programs unless they are represented frequently in the dataset.
%



 
In Figure \ref{fig:embedding}, we show an example of abstract dataflow embedding for an example $d_2$ in Figure 1: \texttt{str = malloc(10 * argc)}. This definition used an API call, {\tt malloc}, with the constant {\tt 10}, operator {\tt *}, and data type {\tt char*}. Contrasted with the 3-bit bit vector  (the example in Figure 1 includes three variable definitions) that represents a concrete definition in dataflow analysis for this program, the abstract embedding is larger but a fixed size, consisting of 5x4 elements for this example. Here, 4 is the four properties we considered and 5 is the hyper-parameter $k$ we mentioned above, which defines the size of the pre-defined dictionary, and the length 5  hot-vector encoding represents the value of the property. Because the vector that encodes the abstract dataflow embedding has a fixed size, our embedding approach can scale to any program size in the dataset without impacting the model's efficiency.
The vectors in different programs encode common properties of definitions, so the model can capture the dataflow patterns across programs.

Abstraction potentially brings in approximation. Using the abstract dataflow embedding, two different definitions may lead to the same encoding.
The embedding is designed to be sparse enough that within a program, unique definitions are often represented by unique embedding keys, which allows the model to distinguish definitions within the same function, similar to the bit-vector used in dataflow analysis.

\subsection{Using Graph Learning to Propagate Dataflow Information
}~\label{sec:dataflow-walkthrough}





Our goal in utilizing graph learning is to learn a node embedding that contains dataflow information. Without loss of generality, we use {\it reaching definition} as an instance of dataflow analysis for our explanations.  Our approach takes the following steps. First, we construct the CFG for a program. Second, we perform static analysis to identify all the definitions in the CFG. We then initialize each node of the CFG using the abstract dataflow embedding, based on whether the node is a definition or not. The abstract dataflow embedding is computed from all the programs in the training dataset (see \S \ref{sec:abstract-embedding} for details).

Once the nodes are initialized, we apply the message-passing algorithm~\cite{gilmerNeuralMessagePassing2017a} from graph learning to propagate the dataflow information throughout the CFG, similar to {\it Kildall's method}~\citep{kildall_unified_1973}.
The main differences are that (1) we propagate the abstract dataflow embeddings of the CFG nodes, and (2) instead of using the dataflow equations of transfer function and meet operator, we alternatively apply the {\aggregate} and {\update} functions defined in Equations~\ref{eq:aggregate} and \ref{eq:update} (See \S 3.3).
Although the analogy applies for all GNN architectures trained with message-passing, we implemented our approach using a {\it Gated Graph Sequence Neural Network (GGNN)}~\citep{li_gated_2016}, where {\aggregate} is an {\it Multi-Layer Perceptron (MLP)} and {\update} is a {\it Gated Recurrent Unit (GRU)}; we will use this architecture as an example to compare the two algorithms.

When dataflow information arrives at the merge point of a branch in CFG, graph learning applies the {\aggregate} function. Specifically, in GGNN, the MLP calculates a weighted sum of the representations of multiple neighboring predecessors, resulting in a single vector; this fulfills the same function as the meet operator. When dataflow information arrives at a new node, the {\update} function in graph learning computes the next state by combining the information in the current node with the output of  {\aggregate} from its predecessors. Specifically, in GGNN, the GRU selectively forgets portions of the previous state and integrates new information from the current node and from the neighboring states, similar to the set union/difference with GEN/KILL performed in the transfer function.
Through applying {\aggregate} and {\update}, the initial embedding will be updated with the dataflow information from the neighboring nodes, similar to the effect of dataflow analysis.

As \citet{cummins_programl_2021} noted, DFA iterates to a fixpoint and thus propagates information throughout the entire graph, while graph learning performs a fixed number of iterations $t$ and thus propagates to neighbors in a distance $t$. We set $t$ to the setting which maximized validation-set performance. 

Finally, we combine the learned abstract node embeddings to produce graph level representation using {\it Global Attention Pooling}~\cite{li_gated_2016}, and pass it to a classifier to predict the function as vulnerable or non-vulnerable.



The {\aggregate} and {\update} functions are learned from labeled data during training, rather than using a fixed formula as in dataflow analysis.
By learning from data, we provide an alternative solution to the challenges that often block dataflow analysis such as tracking pointers and handling library calls. Importantly, we no longer need to explicitly specify vulnerability conditions, as required in static analysis. Through learning from training examples, the classifier can capture patterns of dataflow information that represent various types of vulnerabilities and also select the relevant dataflow information for vulnerability detection.

{
In Table \ref{fig:dataflow-step-through}, we step through a reaching definition analysis for the CFG example in Figure \ref{fig:overview}
to demonstrate how dataflow information propagates through the graph and how our approach uses dataflow information for vulnerability detection.



\begin{table}[t]
\centering
\caption{$OUT[v]$ at each iteration of DFA}\label{fig:dataflow-step-through}
\begin{tabular}{ccccc}
\toprule
Iteration & $v_1$ & $v_2$ & $v_3$ & $v_4$ \\
\midrule
0 & $\irow{0 & 0 & 0}$ & $\irow{0 & 0 & 0}$ & $\irow{0 & 0 & 0}$ & $\irow{0 & 0 & 0}$ \\
1 & $\cirow{1 & 0 & 0}$ & $\irow{0 & 0 & 0}$ & $\cirow{0 & 1 & 0}$ & $\cirow{0 & 0 & 1}$ \\
2 & $\irow{1 & 0 & 0}$ & $\cirow{1 & 0 & 0}$ & $\irow{0 & 1 & 0}$ & $\cirow{0 & 1 & 1}$ \\
3 & $\irow{1 & 0 & 0}$ & $\irow{1 & 0 & 0}$ & $\irow{0 & 1 & 0}$ & $\cirow{1 & 1 & 1}$ \\
\bottomrule
\end{tabular}
\end{table}

The row {\it Iteration 0} shows the initialization of each node in the reaching definition analysis. At iteration 1, the DFA updates $OUT[v_1]$, $OUT[v_3]$ and $OUT[v_4]$ using the transfer function to indicate that the new definitions are introduced at the nodes. At iteration 2, $OUT[v_1]$ (including $d_1$) propagates to $v_2$ and $OUT[v_3]$ (including $d_2$) propagates to $v_4$, through the CFG edges. At iteration 3, the meet operator is used to combine $OUT[v_2]$ and $OUT[v_3]$. Specifically, $IN[v_4] = \bigcup \{ OUT[v_2], OUT[v_3] \}$, computed as $\irow{1 &0 & 0} \lor \irow{0 &1 &0} =  \irow{1& 1 &0}$; then the transfer function combines $IN[v_4]$ with $GEN_{v_4}$, resulting in $OUT[v_4] = \irow{1 & 1 & 1}$.

After the DFA algorithm terminates, the final states of the nodes are used to detect vulnerabilities. The state of $v_4$ is $\irow{1 & 1 & 1}$, which indicates that both $d_1$ and $d_2$ may reach $v_4$ depending on the program values. Because the definition $d_1: \texttt{str = NULL}$ can reach the dereference at $v_4$, we can conclude that this program has a null-pointer dereference vulnerability.
Similarly, in graph learning, after a fixed number of iterations,
all the node representations are combined using a graph readout operation to produce a graph-level representation, which is used to predict for vulnerability detection. Programs with the null-pointer dereference bugs will have the same abstract definitions characterized by the {\tt char*} type and the constant {\tt NULL} to reach the pointer dereference statements.
We believe that the dataflow information represented by {\tool} will allow a relatively simple classifier to recognize this pattern among the training dataset.

\begin{removeenv}
\todo{shorten this part}

To perform this analysis, DFA propagates the definitions which are ``generated'' along the control flow of the program and marks a definition as ``killed'' when another definition overwrites the variable.
The domain is the bit vector of all definitions in the program\bent{; here, $\irow{d_1 & d_2}$,} where each bit $d_i \in \{0, 1\}$ represents a definition.
Table \ref{fig:dataflow-step-through} shows each step of the algorithm.
The initial states of all nodes are $\irow{0 & 0}$.
In iteration 1, $v_1$ generates $\irow{1 & 0}$ and $v_3$ generates $\irow{0 & 1}$ because $d_1$ and $d_2$ are generated in $v_1$ and $v_3$ respectively.
In iteration 2, $d_1$ propagates to $v_2$ along the CFG edge $v_1 \rightarrow v_2$; same for $d_2$ to $v_3$. $d_2$ reaches $v_3$, but $d_1$ is killed (overwritten) and so $OUT(v_3) = \irow{0 & 1}$.
In iteration 3, the meet operator is used; specifically, $IN(v_4) = \bigcup \{ OUT(v_2), OUT(v_3) \} = \{ \irow{1 & 0}, \irow{0 & 1} \} = \irow{1 & 1}$.
$\bigcup$ is used as the meet operator so that $v_4$ incorporates all possible definitions which may reach it depending on the concrete program values.
Finally, $\irow{1 & 1}$ is propagated to $v_4$, which indicates that both $d_1$ and $d_2$ may reach $v_4$ depending on the program values. Because the definition $d_1: \texttt{str = NULL}$ can reach the dereference at $v_4$, we can conclude that this program path has a potential null-pointer dereference bug.

One minor difference between the two algorithms is that DFA iterates until it reaches a fixpoint, while GNN performs a fixed number of iterations.
Kam et al.~\citet{kam_monotone_1977} showed that DFA converges within $d(G) + 3$ passes, where $d$ is the loop connectedness of $G$.
We set the hyperparameter $t$ based on the best validation performance through our experimentation (see Section \ref{sec:training-appendix}). \ben{Note, we did not rigorously set $t$ based on the best validation performance - we only set a value 1000 based on experimentation with values 10, 100, 1000, 5000, 10000.}
\end{removeenv}

\newcommand{\mysubsection}[1]{\noindent{\textbf{#1:}}}
\newcommand{\combinedtool}{DeepDFA+LineVul }
\newcommand{\entry}[2]{\makecell[ct]{#1\\(#2)}}
\newcommand{\entryoneline}[2]{\makecell[ct]{#1\\\wip{(#2)}}}


\section{Evaluation}\label{sec:experiments}
In the evaluation, we studied 3 research questions: 
\begin{enumerate}
    \item Is {\tool} \textbf{effective} for finding vulnerabilities? 
    \item Is {\tool} \textbf{efficient}, both in terms of training data and computational resources?
    \item Can {\tool} \textbf{generalize} to unseen projects?
\end{enumerate}

We also performed ablations on \tool{} to understand the effects of each feature on its performance.

\subsection{Implementation}\label{sec:experiment-setup}
To explore whether DeepDFA can advance the state-of-the-art, we created two settings, {\bf \tool} and {\bf DeepDFA+LLM}.
We implemented {\tool} using the GGNN architecture~\citep{li_gated_2016} and based on LineVD's implementation\footnote{\url{https://github.com/davidhin/linevd}}, using PyTorch and DGL\footnote{All of our code and data are available at \url{https://doi.org/10.6084/m9.figshare.21225413}}. We used Joern\footnote{Joern version 1.1.1072, available at \url{https://joern.io}} to parse the CFGs because it does not require compilation; this allows our approach to be utilized out-of-the-box given only the source code, without extra configuration.
To implement DeepDFA+LLM, during training and inference, we combine the graph embedding generated by DeepDFA's graph readout stage with the sentence embedding produced by the final self-attention layer of LLM. The embeddings are concatenated and fed into a feed-forward classifier layer; both embeddings and the classifier are trained jointly.
We believe that providing dataflow information can improve the LLM embedding, as LLM is trained exclusively from text and it is hard to learn dataflow relations among all the dependencies of tokens.

To avoid data leakage, we extracted the initial abstract dataflow embedding from the training set only. We set the hyperparameters $k$ and $t$ based on the best validation performance through our experimentation. When $k = 1000$, the model covered most (79.38\%) of the definitions in the test dataset.
That means, the dictionary still misses some APIs, constants, data types or operators that occur in the test data set but are not frequent in the training dataset. To learn a more general representation and improve the coverage for test dataset, in future work,
we can train the abstract dataflow embedding using a very large dataset of code, e.g., using self-supervised learning without  the need of vulnerability labels.




\begin{table}[t]
    \centering
    \caption{Hyperparameters used for training \tool.
    }
    \label{fig:hyperparams}
    \begin{tabular}{lr}
        \toprule
        Hyperparameter & Value \\
        \midrule
        $\lambda$ (learning rate) & $1e^{-3}$ \\
        $L_2$ weight & $1e^{-2}$ \\
        $k$ (threshold) & 1000 \\
        $t$ (number of GNN steps) & 5 \\
        Hidden size & 32 \\
        \# output layers & 3 \\
        Batch size & 256 \\
        \bottomrule
    \end{tabular}
\end{table}

For reproducibility, Table \ref{fig:hyperparams} documents the hyperparameters we used for training \tool{}.

\subsection{Experimental setup}


%
%
%
%
%
In the recent literature~\citep{fu_linevul_2022,li_vulnerability_2021,chakraborty_deep_2022}, vulnerability detection models are typically evaluated with the {Devign}~\citep{zhou_devign_2019} or {\dataset}~\citep{fan_cc_2020} datasets, both of which contain real-world open-source C/C++ projects.
In our evaluation, we used the {\dataset} dataset because
(1) it is bigger than Devign, consisting of 188,636 functions with 10,900 (6\%) vulnerable labels and 177,736 (94\%) non-vulnerable labels, and
(2) it reflects the imbalanced distribution of real-world code (Devign is a balanced dataset),
with the minority of code being labeled as vulnerable~\citep{chakraborty_deep_2022}.
To corroborate our results, we also evaluated the models on \textsc{DbgBench}~\cite{dbgbench}, explained further in RQ3.

\mysubsection{Scope} Currently, DeepDFA reports whether a function is vulnerable or not. We can further apply deep learning explanation tools to report line level vulnerabilities, as done in Li et al.'s work~\cite{li_vulnerability_2021}. We leave this evaluation for future work. We evaluated \tool{} on C/C++ programs, as done by the most deep learning-based vulnerability detection tools. However, we believe that DeepDFA can also be applied to other popular programming languages, such as Python and Java. This is because we extract the abstract dataflow embedding (API, datatype, literal, operator) from the training dataset, independent of the programming language.



\mysubsection{RQ1}
To evaluate the models' performance, we trained the models on the train/validation/test splits of 80/10/10\% published by the LineVul paper~\citep{fu_linevul_2022}.
 To address class imbalance while training {\tool}, we undersampled the majority class (non-vulnerable) following Japkowicz et al.~\cite{japkowicz2000class}; our initial studies found that this improved our performance on the validation set.
 We kept the original ratio of vulnerable/non-vulnerable labels for the validation and test sets.
We used Joern\footnote{\url{https://joern.io}}~\citep{yamaguchi_modeling_2014} to parse the code into its CFG representation.
Joern could not parse some programs in the dataset (0.8\%; see Appendix \ref{sec:appendix-joern-failures} for details), so we used the remaining data in our experiments.
The performance of the baseline models was similar to the full dataset (see Appendices \ref{sec:appendix-100-1} \& \ref{sec:appendix-100-2} for details).

We report the following performance metrics:


\begin{itemize}
    \item \textbf{Precision} reports the portion of positive predictions which were correct: $P = \frac{TP}{TP+FP}$.
    \item \textbf{Recall} calculates the portion of positive examples which were recalled correctly: $R = \frac{TP}{TP+FN}$.
    \item \textbf{F1} is the harmonic mean between Precision and Recall: $F1 = 2 * \frac{P * R}{P + R}$. We used $F1$ to decide the highest performing model because it balances precision and recall, which are both important in an imbalanced dataset.
\end{itemize}

Since the model performance can vary with different random seeds~\cite{empiricalstudy}, we trained the models 3 times with different random seeds and reported the mean score and standard deviation for each metric.
We used McNemar's statistical test, following best practices~\cite{dietterich_approximate_1998}, to confirm that our 
improvement is statistically significant, using the implementation in \textit{statsmodels} v0.14.0~\cite{seabold2010statsmodels}.


\mysubsection{RQ2}
To evaluate the models' efficiency in terms of computational resources, we measured the runtime and memory usage. These are often contested resources in deep learning workloads~\cite{howard_mobilenets_2017}.
We report the following metrics for runtime and memory usage:
\begin{itemize}
    \item \textbf{Training time}: the wall-clock time to execute one training run with one validation run per epoch
    \item \textbf{Inference time}: the average wall-clock time to predict for one example.
    \item \textbf{MACs}: the average number of Multiply-Accumulate operations\footnote{We used DeepSpeed profiler to measure MACs~\cite{deepspeed}. \url{https://www.deepspeed.ai}} to predict for one example; this measures the performance independently of the computing platform~\citep{tan_efficientnet_2020,howard_mobilenets_2017}.
    \item \textbf{Parameter count}: the number of trainable parameters in the neural network model.
\end{itemize}

We evaluated the runtime on an AMD Ryzen 5 1600 3.2 GHz processor with 48GB of RAM and an Nvidia 3090 GPU with 24GB of GPU memory.


To evaluate the models' efficiency in terms of training data, we trained the models on progressively smaller subsets which we randomly sampled from {\dataset} (100\%, 10\%, 1\%, 0.5\%, 0.1\%, shown in the columns of Table \ref{fig:dataset-sizes}). Each subset includes the smaller subsets (e.g. 10\% subset includes the 1\% subset and 1\% includes 0.5\%). We generated 3 versions of the subsets using different random seeds and reported the mean and standard deviation F1 score. The goal of this study is to discover what are the minimum training data needed for these models to perform well on the test dataset.





\mysubsection{RQ3} We prepared two experiments for this RQ. In the first experiment, we created  a dataset from {\dataset} to evaluate how well the models generalize to unseen projects. This dataset consists of the {\it mixed-project} and {\it cross-project} two settings.
To set up the mixed-project setting, we held out 10k randomly selected examples for the validation and test sets and used the rest for training, similar to the original method of partitioning the dataset. The training set and test set can and often do contain examples from the same project, though individual examples will not be duplicated between the two sets.
To set up the cross-project setting, we held out 10k examples from randomly selected projects in {\dataset} for the validation and test sets, and used the rest of the projects for training. The projects in the test set are distinct from the projects in the training set.
To mitigate the potential bias caused by the selection of projects, we repeated this process 5 times with different selections of the cross-project data and report the results of 5-fold cross validation.

In order to further evaluate generalization to unseen projects, we applied DeepDFA on buggy and patched programs from \textsc{DbgBench}~\cite{dbgbench}.
\textsc{DbgBench} consists of a set of real-world C programs with bugs, which were analyzed and fixed by professional software engineers. The \textsc{DbgBench} programs are distinct from the programs in the {\dataset} dataset.
We labeled the buggy functions using the fault locations documented in \textsc{DbgBench}; these were labeled by the consensus of multiple developers and were manually checked for correctness, and thus are more reliable than {\dataset}'s labeling process based on bug-fixing commits.
We included all functions which had a bug location marked as ``buggy'' and their corresponding patched versions in our study.
We excluded the bugs marked as ``Functional'' because these bugs cannot be detected without program-specific bug constraints.
We only included the patched versions which were modified by the developers' fixes, taking the first correct\footnote{Marked in \textsc{DbgBench} as ``Developer fix'' or ``Different but Correct Fix'', e.g. \url{https://github.com/dbgbench/dbgbench.github.io/blob/master/patches/find.dbcb10e9/README.md}} developer patch which could be applied to the program. We included only one patch to reduce the effects of code duplication, which can unfairly bias test performance~\cite{allamanis_adverse_2019}. Since the models only view the function-level context, they will not produce a different prediction on the functions which were not modified by the patch. 
We also excluded 8 examples which could not be processed by Joern in order to fairly compare the models' performance scores.
This resulted in a dataset of 22 programs: 17 buggy + 5 patched.
We evaluated the checkpoints trained from 3 random seeds in Section \ref{sec:effectiveness} and report the mean performance scores.

\mysubsection{Ablation study}
%
We ran two ablation settings for each of the four abstract dataflow embedding features, resulting in eight settings: (1) using one feature at a time and (2) using three features at a time (leaving one out).
In each setting, we trained \tool{} on the Big-Vul~\cite{fan_cc_2020} training dataset, and then evaluated on the Big-Vul test dataset and \textsc{DbgBench}~\cite{dbgbench}.

\mysubsection{Baselines}
We compared against \wip{7} non-transformer models: VulDeePecker, SySeVR, Draper, Devign, ReVeal, ReGVD, IVDetect,
and 4 large language models: CodeBERT, LineVul, UniXcoder, and CodeT5
\footnote{We could not reproduce LineVD and ContraFlow on {\dataset} for function-level vulnerability detection.}.
These models were developed recently with diverse architectures, and they represent the state-of-the-art of vulnerability detection models~\cite{empiricalstudy}.
See Section \ref{sec:related-work} for an overview of the models and Appendix \ref{sec:training-appendix} in the supplementary materials for the details of our reproductions.





\subsection{Effectiveness}\label{sec:effectiveness}

\begin{table}[t]
    \caption{{\tool} outperformed the baselines and can be used to further improve the existing model performance.
    All scores are reported as \textit{Mean (Standard deviation)}.
    Note that VulDeePecker, SySeVR, Draper, and IVDetect performance were directly taken from the IVDetect paper~\citep{li_vulnerability_2021}, so we do not report the variance.
    }
    \label{fig:compare-performance}
    \begin{subtable}[t]{.5\textwidth}
    \centering
    \caption{Comparison with non-transformer models.
    }\label{fig:compare-nontransformer}
    \begin{tabular}{lccc}
    \toprule
    Model & F1 & Precision & Recall \\
    \midrule
    VulDeePecker	& 12.00	& 49.00	& 19.00 \\
    \mytablespace
    SySeVR	        & 15.00	& {\bf 74.00}	& 27.00 \\
    \mytablespace
    Draper	        & 16.00	& 48.00	& 24.00 \\
    \mytablespace
    ReGVD	& \entry{19.15}{2.65} & \entry{63.67}{4.43} & \entry{11.33}{1.94} \\
    \mytablespace
    IVDetect	& 23.00	& 72.00	& 35.00 \\
    \mytablespace
    Devign & \entry{26.85}{0.97} & \entry{29.00}{0.38} & \entry{25.03}{1.67} \\
    \mytablespace
    ReVeal & \entry{32.94}{0.75} & \entry{34.27}{1.58} & \entry{31.73}{0.65} \\
    \mytablespace
    
    {\tool}             & \entry{\textbf{68.26}}{0.16}	& \entry{53.98}{0.06}	& \entry{\textbf{92.81}}{0.40} \\
    \bottomrule
    \end{tabular}
    \end{subtable}
    %
    \newline
    \vspace*{0.1 cm}
    \newline
    \begin{subtable}[t]{.5\textwidth}
    \centering
    \caption{Comparison with transformer models.
    }\label{fig:compare-transformer}
    \begin{tabular}{lcccrr}
    \toprule
    Model & F1 & Precision & Recall \\
    \midrule
    CodeBERT	        & \entry{21.04}{6.72} & \entry{68.48}{11.76} & \entry{12.91}{5.51} \\
    \mytablespace
    CodeT5             & \entry{45.61}{0.71} & \entry{56.47}{6.22} & \entry{38.56}{2.45} \\
    \mytablespace
    LineVul             & \entry{93.23}{0.31} & \entry{97.32}{0.66} & \entry{89.48}{0.42} \\
    \mytablespace
    UniXcoder             & \entry{95.11}{0.21} & \entry{96.96}{1.14} & \entry{93.34}{1.23} \\
    \mytablespace
    
    {\tool}             & \entry{68.26}{0.16}	& \entry{53.98}{0.06}	& \entry{92.81}{0.40} \\
    \mytablespace
     {{\tool}+CodeT5}     & \entry{81.39}{0.96} & \entry{94.23}{2.98} & \entry{71.67}{1.09} \\
    \mytablespace
    {\combinedtool}     & \entry{96.40}{0.13} & \entry{\textbf{98.69}}{0.28} & \entry{94.22}{0.46} \\
    \mytablespace
    {{\tool}+UniXcoder}     & \entry{\textbf{96.46}}{0.09} & \entry{97.82}{0.99} & \entry{\textbf{95.14}}{1.09} \\
    \bottomrule
    \end{tabular}
    \end{subtable}
\end{table}

\mysubsection{Comparison with non-transformer models}
In Table \ref{fig:compare-nontransformer}, we show that {\tool} performed much better than the baseline models on F1 score and recall. {\tool}'s score was 47.51 higher than the average F1 score computed over all the baselines. In addition,
 compared to the other 6 models, {\tool} reported lower variances for all the three metrics.
This indicates that {\tool} was more robust to random noise throughout training, and thus more likely to perform as expected after training.


%
The results show that our abstract dataflow embedding indeed encodes useful information for vulnerability detection, despite the fact that the node representation is small and the graph is simple. It is more effective than {\it property graphs} (a combination of AST, CFG, and PDG) used in Devign and Reveal. These baseline models represented nodes using unsupervised word embeddings~\cite{mikolov_efficient_2013,le_distributed_2014,pennington_glove_2014}, which do not have a direct relationship with vulnerabilities.
In contrast, {\tool}'s node representation encodes the dataflow sets of reaching definitions, related to the root causes of vulnerabilities.






\mysubsection{Comparison with transformer models}
We compared \tool{} with CodeBERT, LineVul, UniXcoder, and CodeT5 -- the state-of-the-art among the transformer language models we evaluated.
(see Appendix~\ref{sec:appendix-100-1} for the performances of all the baseline models).
Table \ref{fig:compare-transformer} shows that \tool{} performed considerably better than CodeBERT and CodeT5 in F1 score and had the smallest variance (among all the models) between runs. 


Although UniXcoder and LineVul performed better than \tool{} in terms of F1 score, \tool{}'s embedding can be combined with UniXcoder and LineVul to further improve their performance.  We achieved state-of-the-art performance on all three metrics by adding \tool{}'s embedding to UniXcoder, with an F1 score of 96.46 (1.35 improvement), a Precision score of 97.82 (0.86 improvement), and a Recall score of 95.14 (1.80 improvement).
Adding \tool{}'s embedding improved CodeT5 considerably -- by 35.78 F1 score -- and improved LineVul by 3.17 F1 score.
We used McNemar's significance test, as recommended by Dietterich et al.~\cite{dietterich_approximate_1998}, to confirm that the differences in performance were statistically significant
($p < 0.05$; see Table \ref{fig:statistical-test}).

\begin{table}[t]
    \centering
    \caption{Results of statistical tests for model comparison.}
    \label{fig:statistical-test}
    \begin{tabular}{lrr}
    \toprule
    Models compared                 & $\chi^2$ statistic & $p$-value \\
    \midrule
    LineVul vs. \tool{}+LineVul        & 20.0 & \num{2.77e-07} \\
    UniXcoder vs. \tool{}+UniXcoder    & 25.0 & \num{5.35e-4} \\
    \bottomrule
    \end{tabular}
\end{table}

{\tool} does not use any text-/token-level information such as variable and function names, yet it has achieved excellent performance. We believe that leveraging the domain-specific algorithm of reaching definition analysis to guide graph learning indeed plays an important role and that the embedding indeed encodes semantic features (e.g., data relations) that are important for vulnerability detection. The fact that {\tool} can further improve the top-performing LLMs indicates that LLMs, which exclusively leverage text information, may not sufficiently learn the dataflow of code; \tool{} thus provides the complementary information for vulnerability detection.
We further believe that the examples which {\tool} predicted incorrectly could be attributed to the fact that reaching definition analysis cannot handle all types of vulnerabilities. Thus, by adding other dataflow analyses such as live variable analysis, \tool{} could further improve its performance. We will leave such an investigation to our future work.





\begin{tcolorbox}
\textbf{RQ1 result:} {\tool} performed much better than all non-transformer baselines.
When combined with transformer models, it achieved the highest SOTA score on all metrics.
\end{tcolorbox}











\subsection{Efficiency}
\label{sec:efficiency}

\mysubsection{Efficiency of computational resources}
In Table \ref{runtime}, we present the runtime comparison of {\tool}, LineVul, and UniXcoder. Here, we did not list other models because their performances are much worse (shown in Table \ref{fig:compare-performance}), and they took hours to train (see Appendix \ref{sec:all-training-times} in the supplementary material), compared to {\tool} which finished training in \wip{9 minutes (excluding data preprocessing time)}. In Appendix \ref{sec:model-sizes}, we also listed the sizes of the models in terms of the number of parameters.

\begin{table}[t]
\centering
\caption{{\tool}'s training/inference time was faster than the baselines.}\label{runtime}
\begin{tabular}{lrrrr}
\toprule
& \multirow[b]{2}{*}[1pt]{\makecell[c]{Train time\\(ms)}} & \multicolumn{3}{c}{Inference cost per example} \\
\cmidrule{3-5}
Model & & \makecell[r]{GPU (ms)} & \makecell[r]{CPU (ms)} & \makecell[r]{MACs}\\
\midrule
LineVul         & 10h19m  & 11.1 & 1068.2 & 48.32 B  \\
{DDFA+LV} & 10h40m  & 15.4 & 1571.5 & 48.32 B  \\ 
UniXcoder & 11h16m & 9.5 & 486.0 & 48.32 B  \\
DDFA+UXC & 13h22m & 13.3 & 922.1 & 48.32 B  \\
{\tool}         & \textbf{9m}     & \textbf{4.6} & \textbf{5.8} & \textbf{40.27 M}      \\
\bottomrule
\end{tabular}
\end{table}



Compared to UniXcoder, DeepDFA took 75x less time to train, 2x faster inference on GPU, and 84x faster inference on CPU.
{\tool} had the least parameters of all models, equal to 67\% of the smallest model (ReVeal) and 0.3\% of the highest-performing baseline model (UniXcoder).
These results consistently indicate that {\tool} excels in its efficiency compared to other models. This is possible because {\tool} is based on the dataflow analysis's compact representation -- bitvector, which captures the relevant semantic information in bits and thus is more efficient compared to tokenized strings. {\tool} propagated information along only the domain-specific CFG edges, rather than associating every pair of tokens in an exhaustive fashion.

DeepDFA's short inference time due to a low number of MAC operations enables its use in non-GPU environments (which are common for software development) where large language models may not be easily deployed. DeepDFA's short training time enables techniques like per-project fine-tuning and hyperparameter tuning, which would be much more costly with the LLMs' training times of over 10 hours.
Because of {\tool}'s small parameter count, it is ideal for resource-limited computing platforms such as mobile devices, where large models cannot be used~\citep{howard_mobilenets_2017}.


\mysubsection{Efficiency on training data}
%
In Table \ref{fig:dataset-sizes}, we report the performance of {\tool} over reduced training dataset sizes, compared to the SOTA models, LineVul and UniXcoder.
The columns ``\# data'' and ``\# vul''  list the number of training examples in each subset and, of these, the number of vulnerable examples.
The results show that {\tool} maintained a stable performance across small dataset sizes, even with only 0.1\% of the training dataset, using only 151 training examples. In contrast, LineVul and UniXcoder steadily dropped in performance as the size of the training dataset decreased. At 0.1\% data, LineVul's mean performance was only 29.75 in F1 and UniXcoder's was 4.36. 

For project-specific training in applications within a single development team, a model which can learn efficiently from a small dataset is useful. 




\newcommand{\dataentry}[2]{#1}

\begin{table}[t]
\caption{{\tool} retained its performance on limited data.}
\label{fig:dataset-sizes}
\begin{tabular}{rrrrrr}
\toprule
&&&\multicolumn{3}{c}{F1}\\ \cmidrule{4-6}
Portion & \# data    & \# vul    & LineVul & UniXcoder & \tool{}     \\
\midrule

0.1\%   & 151     & 11    & \dataentry{29.75}{31.68} & \dataentry{4.36}{5.15} & \dataentry{\textbf{55.24}}{22.47} \\
0.5\%   & 755     & 56    & \dataentry{77.69}{2.70}  & \dataentry{\textbf{79.37}}{4.74} & \dataentry{68.17}{0.41}         \\
1.0\%     & 1,510   & 90    & \dataentry{\textbf{84.62}}{1.21}  & \dataentry{79.60}{6.10} & \dataentry{68.40}{0.0}       \\
10.0\%    & 15,091  & 885   & \dataentry{86.67}{0.14}  & \dataentry{\textbf{92.53}}{0.53} & \dataentry{68.44}{0.07}         \\
100.0\%   & 150,908 & 8,736 & \dataentry{93.23}{0.31}  & \dataentry{\textbf{95.11}}{0.21} & \dataentry{68.26}{0.16}  \\
\bottomrule


\end{tabular}
\end{table}

We believe that our model's stable performance over the reduced dataset and good performance with very small training datasets demonstrate the advantage of the small models and the effectiveness of domain-specific algorithms to guide model learning.
{\tool} is less prone to overfitting to datasets of limited size since it has fewer parameters than LineVul~\cite{bishop:2006:prml}.
On the other hand, the transformer models require a large corpus of programs to learn the patterns among the unstructured token data. 



\begin{tcolorbox}
\textbf{RQ2 result:} {\tool} was considerably faster than the baselines; it took 9 minutes to train, 4.64 milliseconds for inference on GPU, and 5.8 milliseconds for inference on CPU.
{\tool} retained stable performance as the training dataset size was reduced.
In a low-data scenario, \tool{} outperformed LineVul and UniXcoder by 25.49 and 50.88 points F1 score.
\end{tcolorbox}

\subsection{Generalization}\label{sec:generalization}

\begin{table}[b]
    \centering
    \caption{How do the models handle unseen projects? Note the performance drop ($\Delta$F1) from the cross-project to mixed-project setting.}
    \label{fig:cross-project-drop}
    \begin{tabular}{lrrr}
\toprule
Model               & Mixed F1 & Cross F1     & $\Delta$F1      \\
\midrule
LineVul           & 84.03     & 71.37       & -12.66   \\

UniXcoder & 86.30 & 76.72 & -9.58 \\
{\tool}           & 70.49     & 68.58     & \textbf{-1.91}     \\

{\wip{DeepDFA+LineVul}} & 87.89      & 71.88     & -16.02   \\

DeepDFA+UniXcoder & \textbf{89.85} & \textbf{78.07} & -11.77 \\
\bottomrule
    \end{tabular}
\end{table}


\mysubsection{Cross-project evaluation on {\dataset}}
We compared the models' F1 scores on the {\it cross-project} (shown as Cross F1) and {\it mixed-project} (shown as Mixed F1) settings to evaluate the models' capabilities of generalizing over unseen projects.
Table \ref{fig:cross-project-drop} presents the highest-performing baseline models, LineVul and UniXcoder, compared to DeepDFA (the results of the other baseline models are available in the supplementary material, Appendix \ref{sec:appendix-100-2}).
Among the most important metrics, $\Delta$F1 shows how performance changes when the model is applied to unseen projects. Shown under Column $\Delta$F1,  {\tool} only dropped 1.91 (2.7\%) F1 score, compared to 12.66 (15.1\%) drop for LineVul and 9.58 (11.1\%) drop for UniXcoder.
{\tool{}+UniXcoder} reported the best performance for both the mixed-project and cross-project settings, improving on UniXcoder's mean F1 score by 3.55 and 1.35 points respectively;
{\tool{}+LineVul} also improved LineVul's F1 score in both settings.


\noindent{\bf Applying to \textsc{DbgBench}:}
%
%

In Table \ref{fig:dbgbench}, we report our experience of applying deep learning tools to real-world bug benchmarks.
{\tool} detected 8.7 out of 17 total bugs on average across 3 runs. {\tool} also correctly predicted non-vulnerable for 3 out of 5 patched programs.
On the other hand, neither of the competing LLMs, LineVul and UniXcoder, detected any bugs and in fact both models reported all programs as non-vulnerable with high confidence. This implies that these models were heavily biased to predict all examples in \textsc{DbgBench} as non-vulnerable.
With the addition of \tool{}, {\tool{}+LineVul} and {\tool{}+UniXcoder}'s generalization greatly improved, yet they did not perform as well overall as \tool{} alone.
It should be noted that the bugs in \textsc{DbgBench} are very complex and took human experts hours to diagnose~\cite{dbgbench}.
In the past, we have tried a variety of static analysis tools, such as Cppcheck\footnote{\url{https://cppcheck.sourceforge.io/}} and Polyspace\footnote{\url{https://www.mathworks.com/products/polyspace.html}}, to detect bugs in \textsc{DbgBench}, but we have not detected any of these bugs.




We believe that {\tool} generalizes better because it does not rely on spurious features that may exist at token and text level, such as variable names and function names, as reported by previous research~\citep{chakraborty_deep_2022}. 
These spurious features are no longer correlated with vulnerabilities in unseen projects, as their input tokens will likely change. Our abstract dataflow embedding encodes the usage patterns of commonly used
API calls, operators, constants, and data types. Such patterns can be extracted from unseen text and are directly related to the cause of the vulnerabilities, and thus might help DeepDFA generalize better over unseen projects.



\begin{tcolorbox}
\textbf{RQ3 result:}
{\tool} had the smallest drop in F1 score ($\Delta$ F1) when applying to the vulnerabilities in the projects that are not seen in training datasets.
{\tool} was able to detect complex bugs in \textsc{DbgBench} and was able to distinguish the buggy and patched versions.
The SOTA models, LineVul and UniXcoder, did not detect any bugs in \textsc{DbgBench}.
\end{tcolorbox}

\begin{table}[t]
    \centering
    \caption{DeepDFA generalized to real-world bugs in \textsc{DbgBench}. Results are averaged over checkpoints from 3 random seeds. Buggy/Patched columns show the number of correct predictions on buggy/patched programs respectively.}
    \label{fig:dbgbench}

    \begin{tabular}{lrrrr}
    \toprule
    Model   & Buggy & Patched & Accuracy & F1 \\
    \midrule
    LineVul & 0.0     & \textbf{5.0}       & 22.73    & 0.00  \\
    \tool{}+LineVul   & \textbf{9.0}   & 1.3    & 46.97    & 60.67  \\
    UniXcoder & 0.0 & 5.0 & 22.73 & 0.00  \\
    \tool{}+UniXcoder & \textbf{9.0}   & 1.3    & 46.97    & 60.67  \\
    \tool{} & 8.7   & 3.0       & \textbf{53.03}    & \textbf{64.29}  \\
    \midrule
    Total   & 17.0    & 5.0       & -\ \ \ \       & -\ \ \ \   \\
    \toprule
    \end{tabular}
\end{table}

\subsection{Ablation studies}\label{sec:appendix-ablation}

\begin{table}[t]
    \centering
    \caption{Ablation study evaluated on \textsc{DbgBench}}\label{fig:ablation-dbgbench}
\begin{tabular}{lrrrr}
\toprule
Feature set & Buggy & Patched & Acc & F1 \\
\midrule
\tool{} & \textbf{8.7}   & \textbf{3.0}       & \textbf{53.03}    & \textbf{64.29}  \\
\midrule
API only & 7.7 & 3.0 & 48.48 & 57.50 \\
Datatype only & 8.0 & 3.0 & 50.00 & 59.26 \\
Literal only & 7.7 & 3.0 & 48.48 & 57.50 \\
Operator only & 7.7 & 3.0 & 48.48 & 57.50 \\
\midrule
Api+datatype+literal & 8.0 & 3.0 & 50.00 & 59.26 \\
Api+datatype+operator & 8.0 & 3.0 & 50.00 & 59.26 \\
Api+literal+operator & 8.0 & 3.0 & 50.00 & 59.26 \\
Datatype+literal+operator & 8.0 & 3.0 & 50.00 & 59.26 \\
\midrule
Total   & 17.0    & 5.0       & \makecell[c]{-}      & \makecell[c]{-}  \\
\bottomrule
\end{tabular}
\end{table}


Table \ref{fig:ablation-dbgbench} shows the model's performance on \textsc{DbgBench}. The model detected the most bugs when using all four features compared to other ablation settings. When using only one feature at a time, the model consistently missed 1-2 bugs which were detected by \tool{}. %
When using three features at a time (leaving one out), the model still consistently failed to detect 1 bug which was detected by \tool{}.

\begin{table}[t]
    \centering
    \caption{Ablation study evaluated on the Big-Vul test dataset}
    \label{fig:ablation}
    \begin{tabular}{lccc}
\toprule
Feature set                     & F1 & Precision & Recall \\
\midrule
{\tool}             & \entry{{\textbf{68.26}}}{0.16}	& \entry{53.98}{0.06}	& \entry{{\textbf{92.81}}}{0.40} \\
\midrule
Datatype only & \entry{68.04}{0.19} & \entry{53.83}{0.31} & \entry{92.46}{0.30} \\
\mytablespace
Literal only & \entry{62.50}{0.81} & \entry{50.52}{1.83} & \entry{82.46}{7.26} \\
\mytablespace
Operator only & \entry{64.47}{0.33} & \entry{52.82}{1.59} & \entry{82.98}{4.90} \\
\mytablespace
API only & \entry{63.67}{0.45} & \entry{50.66}{2.36} & \entry{86.14}{5.58} \\
\midrule
API + datatype + literal & \entry{68.18}{0.10} & \entry{\textbf{54.06}}{0.13} & \entry{92.28}{0.15} \\
\mytablespace
API + datatype + operator & \entry{68.16}{0.13} & \entry{54.03}{0.18} & \entry{92.28}{0.15} \\
\mytablespace
API + literal + operator & \entry{68.11}{0.14} & \entry{53.98}{0.20} & \entry{92.28}{0.15} \\
\mytablespace
Datatype + literal + operator & \entry{68.12}{0.20} & \entry{{54.04}}{0.11} & \entry{92.11}{0.46} \\
\bottomrule
    \end{tabular}
\end{table}

Table \ref{fig:ablation} shows the model's performance on the Big-Vul test dataset. DeepDFA (integrating all the four features) performed the best out of all configurations. When testing one feature at a time, datatype by itself performed better than the other 3 features alone. When we used the combined feature sets, the model performed better than using only one feature.

\section{Threats to Validity and Discussions}
\noindent{\bf Threats:}
We evaluated performance primarily on the {\dataset} dataset because this dataset was supported by all the baseline models. Compared to the Devign dataset, {\dataset} is imbalanced and can better reflect a real-world vulnerability detection scenario.
However, {\dataset}'s data collection process based on bug-fixing commits can introduce label noise and selection bias and as a result, the evaluation could fail to represent real-world performance.
To address the selection bias, we studied settings which reflect more realistic scenarios with reduced training datasets and cross-project generalization;
to address the label noise, we evaluated the models on additional real-world bugs collected in \textsc{DbgBench}, which were labeled by developers and manually checked.


The performances reported in RQ1, RQ2, and RQ3 will be affected by the random noise in the model training and, for RQ2, dataset selection. To mitigate this effect, we generated 3 versions of the subsets using different random seeds and reported the mean performance.

The mixed-project and cross-project performance reported in RQ3 will be affected by the random selection of projects in the training/held-out datasets. To mitigate this effect, we performed 5-fold cross-validation and reported the mean performance.

\noindent{\bf Discussions:}  We believe our approach can be extended to \textit{bit-vector} dataflow problems~\cite{Horwitz:dataflow,repsPreciseInterproceduralDataflow1995}. All problems in this category contain a finite set of dataflow facts and have the same form of transfer functions and meet operators (see Equations \ref{eq:meet-operator} and \ref{eq:transfer-function}).
For example, live variables and available expressions~\cite{repsPreciseInterproceduralDataflow1995} are bit-vector problems that are important for vulnerability detection~\cite{cesare_bugalyzecom_2013}.
We believe a new dataflow analysis can be integrated by: (1) defining an abstract dataflow embedding which can capture the dataflow set of the analysis, (2) configuring the neural network used as the aggregate function in GGNN to better simulate the meet operator (based on whether it is a union or intersection operation), and (3) reversing the CFG edges for backward dataflow problems (as reaching definition is a forward dataflow problem).

\section{Related Work}\label{sec:related-work}






Many works have used GNN for vulnerability detection ~\cite{allamanis_learning_2018,wang_learning_2020,dinella_hoppity_2020,du_leopard_2020,cheng_deepwukong_2021,cao_bgnn4vd_2021,nguyen_regvd_2022}. In several recent approaches,
Devign \citep{zhou_devign_2019}, ReVeal \citep{chakraborty_deep_2022}, IVDetect \citep{li_vulnerability_2021}, and LineVD \citep{hin_linevd_2022} used GNN on program graph representations such as AST, CFG, and PDG, and annotated the nodes with unsupervised or pretrained word embeddings.
The novelty of our work is a bit-vector inspired abstract dataflow embedding based on the analogy of graph learning and DFA algorithms.


Transformer models such as CodeBERT~\citep{feng_codebert_2020}, LineVul~\citep{fu_linevul_2022}, and UniXcoder~\cite{guo_unixcoder_2022} used a token-based program representation pretrained on a large body of NL-PL pairs, and then fine-tuned for vulnerability detection. Using CFGs, our graph learning only propagates the information along semantically important edges instead of trying to learn the relations of each pair of tokens. Thus, our approach is substantially more efficient. Since we have used a semantic-based embedding, we show that we can improve the performance of token based models. The most recent work, ContraFlow \citep{cheng_path-sensitive_2022}, learns embeddings of def-use paths (an output of dataflow analysis), then predicts vulnerability detection using a transformer model. Our work directly emulates dataflow analysis and does not require an expensive pretraining phase.





There were also models that used sequence and CNN architectures.
VulDeePecker \citep{li_vuldeepecker_2018} used BiLSTM on slices considering data dependencies. SySeVR \citep{li_sysevr_2021} used BiGRU on slices and adds data dependencies.
Draper \citep{russell_automated_2019} used CNN and Random Forest.
However, none of these models integrates dataflow analysis in its algorithm.


%


\citet{cummins_programl_2021} formulated dataflow analyses as supervised learning tasks and applied it for device mapping and algorithm classification; we discuss the differences from our work in-depth in Section~\ref{sec:novelty}.
%
%
Other relevant work that explores dataflow analysis and deep learning include: (1) 
\citet{venkatakeerthy_ir2vec_2020} used the output of dataflow analysis, reaching definitions and live variables, to learn flow-aware embeddings; and (2) 
\citet{bielik_learning_2017} and \citet{jeon_machine-learning_2019} learned static analysis formulas from a dataset based on a fixed language. None of these works aims to develop a model for vulnerability detection.



\section{Conclusions and Future Work}
We propose DeepDFA, an efficient graph learning framework and embedding technique for vulnerability detection. Our {\it abstract dataflow embedding} leverages the idea of {\it bit-vector} in dataflow analysis and integrates data usage patterns from semantic features: commonly used API calls, operations, constants, and data types that potentially capture the causes of the vulnerabilities. \tool{} emulates the Kildall method of dataflow analysis using the analogous message-passing algorithm.
Our experimental results show that {\tool} is very efficient.  It is trained in \wip{9 minutes}  and used only 50 vulnerable examples to achieve its top performance. Yet, it still outperformed all non-transformer baselines and generalized the best among all the models.  DeepDFA found bugs in real-world programs from \textsc{DbgBench} while neither of the highest-performing baselines, LineVul and UniXcoder, detected any bugs. Importantly, {\tool} can be used to improve other models.
By combining DeepDFA with the top performing models, we surpassed the state-of-the-art performance for vulnerability detection.

In the future, we plan to incorporate other dataflow analyses, e.g., live variable analysis, that have been used for or vulnerability detection~\citep{cesare_bugalyzecom_2013}.
We also plan to explore the application of explanation tools to precisely pinpoint the vulnerability location at specific lines in the code, and evaluate our framework on detecting vulnerabilities in other programming languages.


\hey{
%
}




\section{Acknowledgements}

We thank the anonymous reviewers for their valuable feedback. This research is partially supported by the U.S. National Science Foundation (NSF) under Awards \#1816352 and \#2313054.

\bibliographystyle{ACM-Reference-Format}
\bibliography{zotero-static,extra}


\begin{thebibliography}{56}


\ifx \showCODEN    \undefined \def \showCODEN     #1{\unskip}     \fi
\ifx \showDOI      \undefined \def \showDOI       #1{#1}\fi
\ifx \showISBNx    \undefined \def \showISBNx     #1{\unskip}     \fi
\ifx \showISBNxiii \undefined \def \showISBNxiii  #1{\unskip}     \fi
\ifx \showISSN     \undefined \def \showISSN      #1{\unskip}     \fi
\ifx \showLCCN     \undefined \def \showLCCN      #1{\unskip}     \fi
\ifx \shownote     \undefined \def \shownote      #1{#1}          \fi
\ifx \showarticletitle \undefined \def \showarticletitle #1{#1}   \fi
\ifx \showURL      \undefined \def \showURL       {\relax}        \fi
\providecommand\bibfield[2]{#2}
\providecommand\bibinfo[2]{#2}
\providecommand\natexlab[1]{#1}
\providecommand\showeprint[2][]{arXiv:#2}

\bibitem[cve(2021)]%
        {cvedetails_2021}
 \bibinfo{year}{2021}\natexlab{}.
\newblock \bibinfo{title}{Browse vulnerabilities by date}.
\newblock
  \bibinfo{howpublished}{\url{https://web.archive.org/web/20211014235218/https://www.cvedetails.com/browse-by-date.php}}.
\newblock
\newblock
\shownote{Accessed November 5 2021}.


\bibitem[ibm(2021)]%
        {ibm_cost_2021}
 \bibinfo{year}{2021}\natexlab{}.
\newblock \bibinfo{title}{{IBM} {Cost} of a {Data} {Breach} {Report} 2021}.
\newblock
\newblock
\newblock
\shownote{Accessed October 29 2021}.


\bibitem[Ahmad et~al\mbox{.}(2021)]%
        {ahmad_unified_2021}
\bibfield{author}{\bibinfo{person}{Wasi~Uddin Ahmad}, \bibinfo{person}{Saikat
  Chakraborty}, \bibinfo{person}{Baishakhi Ray}, {and} \bibinfo{person}{Kai-Wei
  Chang}.} \bibinfo{year}{2021}\natexlab{}.
\newblock \showarticletitle{Unified pre-training for program understanding and
  generation}.
\newblock \bibinfo{journal}{\emph{arXiv preprint arXiv:2103.06333}}
  (\bibinfo{year}{2021}).
\newblock


\bibitem[Aho et~al\mbox{.}(2006)]%
        {aho_compilers_2007}
\bibfield{author}{\bibinfo{person}{Alfred~V. Aho}, \bibinfo{person}{Monica~S.
  Lam}, \bibinfo{person}{Ravi Sethi}, {and} \bibinfo{person}{Jeffrey~D.
  Ullman}.} \bibinfo{year}{2006}\natexlab{}.
\newblock \bibinfo{booktitle}{\emph{Compilers: Principles, Techniques, and
  Tools (2nd Edition)}}.
\newblock \bibinfo{publisher}{Addison-Wesley Longman Publishing Co., Inc.},
  \bibinfo{address}{USA}.
\newblock
\showISBNx{0321486811}


\bibitem[Allamanis(2019)]%
        {allamanis_adverse_2019}
\bibfield{author}{\bibinfo{person}{Miltiadis Allamanis}.}
  \bibinfo{year}{2019}\natexlab{}.
\newblock \showarticletitle{The Adverse Effects of Code Duplication in Machine
  Learning Models of Code}. In \bibinfo{booktitle}{\emph{Proceedings of the
  2019 ACM SIGPLAN International Symposium on New Ideas, New Paradigms, and
  Reflections on Programming and Software}} (Athens, Greece)
  \emph{(\bibinfo{series}{Onward! 2019})}. \bibinfo{publisher}{Association for
  Computing Machinery}, \bibinfo{address}{New York, NY, USA},
  \bibinfo{pages}{143–153}.
\newblock
\showISBNx{9781450369954}
\urldef\tempurl%
\url{https://doi.org/10.1145/3359591.3359735}
\showDOI{\tempurl}


\bibitem[Allamanis et~al\mbox{.}(2018)]%
        {allamanis_learning_2018}
\bibfield{author}{\bibinfo{person}{Miltiadis Allamanis}, \bibinfo{person}{Marc
  Brockschmidt}, {and} \bibinfo{person}{Mahmoud Khademi}.}
  \bibinfo{year}{2018}\natexlab{}.
\newblock \showarticletitle{Learning to Represent Programs with Graphs}. In
  \bibinfo{booktitle}{\emph{International Conference on Learning
  Representations}}.
\newblock
\urldef\tempurl%
\url{https://openreview.net/forum?id=BJOFETxR-}
\showURL{%
\tempurl}


\bibitem[Bielik et~al\mbox{.}(2017)]%
        {bielik_learning_2017}
\bibfield{author}{\bibinfo{person}{Pavol Bielik}, \bibinfo{person}{Veselin
  Raychev}, {and} \bibinfo{person}{Martin Vechev}.}
  \bibinfo{year}{2017}\natexlab{}.
\newblock \showarticletitle{Learning a static analyzer from data}. In
  \bibinfo{booktitle}{\emph{Computer Aided Verification: 29th International
  Conference, CAV 2017, Heidelberg, Germany, July 24-28, 2017, Proceedings,
  Part I 30}}. Springer, \bibinfo{pages}{233--253}.
\newblock


\bibitem[Bishop(2006)]%
        {bishop:2006:prml}
\bibfield{author}{\bibinfo{person}{Christopher~M. Bishop}.}
  \bibinfo{year}{2006}\natexlab{}.
\newblock \bibinfo{booktitle}{\emph{Pattern Recognition and Machine Learning}}.
\newblock \bibinfo{publisher}{Springer}.
\newblock


\bibitem[B\"{o}hme et~al\mbox{.}(2017)]%
        {dbgbench}
\bibfield{author}{\bibinfo{person}{Marcel B\"{o}hme},
  \bibinfo{person}{Ezekiel~Olamide Soremekun}, \bibinfo{person}{Sudipta
  Chattopadhyay}, \bibinfo{person}{Emamurho Ugherughe}, {and}
  \bibinfo{person}{Andreas Zeller}.} \bibinfo{year}{2017}\natexlab{}.
\newblock \showarticletitle{Where is the Bug and How is it Fixed? An Experiment
  with Practitioners}. In \bibinfo{booktitle}{\emph{Proceedings of the 11th
  Joint meeting of the European Software Engineering Conference and the ACM
  SIGSOFT Symposium on the Foundations of Software Engineering}}
  \emph{(\bibinfo{series}{ESEC/FSE 2017})}. \bibinfo{pages}{1--11}.
\newblock


\bibitem[Cao et~al\mbox{.}(2021)]%
        {cao_bgnn4vd_2021}
\bibfield{author}{\bibinfo{person}{Sicong Cao}, \bibinfo{person}{Xiaobing Sun},
  \bibinfo{person}{Lili Bo}, \bibinfo{person}{Ying Wei}, {and}
  \bibinfo{person}{Bin Li}.} \bibinfo{year}{2021}\natexlab{}.
\newblock \showarticletitle{BGNN4VD: Constructing Bidirectional Graph
  Neural-Network for Vulnerability Detection}.
\newblock \bibinfo{journal}{\emph{Inf. Softw. Technol.}} \bibinfo{volume}{136},
  \bibinfo{number}{C} (\bibinfo{date}{aug} \bibinfo{year}{2021}),
  \bibinfo{numpages}{11}~pages.
\newblock
\showISSN{0950-5849}
\urldef\tempurl%
\url{https://doi.org/10.1016/j.infsof.2021.106576}
\showDOI{\tempurl}


\bibitem[Cao et~al\mbox{.}(2022)]%
        {cao_mvd_2022}
\bibfield{author}{\bibinfo{person}{Sicong Cao}, \bibinfo{person}{Xiaobing Sun},
  \bibinfo{person}{Lili Bo}, \bibinfo{person}{Rongxin Wu}, \bibinfo{person}{Bin
  Li}, {and} \bibinfo{person}{Chuanqi Tao}.} \bibinfo{year}{2022}\natexlab{}.
\newblock \showarticletitle{{MVD}: {Memory}-{Related} {Vulnerability}
  {Detection} {Based} on {Flow}-{Sensitive} {Graph} {Neural} {Networks}}.
\newblock \bibinfo{journal}{\emph{arXiv:2203.02660 [cs]}}
  (\bibinfo{date}{March} \bibinfo{year}{2022}).
\newblock
\urldef\tempurl%
\url{https://doi.org/10.1145/3510003.3510219}
\showDOI{\tempurl}
\newblock
\shownote{arXiv: 2203.02660}.


\bibitem[Cesare(2013)]%
        {cesare_bugalyzecom_2013}
\bibfield{author}{\bibinfo{person}{Silvio Cesare}.}
  \bibinfo{year}{2013}\natexlab{}.
\newblock \showarticletitle{Bugalyze.com - {Detecting} {Bugs} {Using}
  {Decompilation} and {Data} {Flow} {Analysis}}. In
  \bibinfo{booktitle}{\emph{{BlackHat} {USA}}}. \bibinfo{pages}{9}.
\newblock


\bibitem[Chakraborty et~al\mbox{.}(2022)]%
        {chakraborty_deep_2022}
\bibfield{author}{\bibinfo{person}{Saikat Chakraborty}, \bibinfo{person}{Rahul
  Krishna}, \bibinfo{person}{Yangruibo Ding}, {and} \bibinfo{person}{Baishakhi
  Ray}.} \bibinfo{year}{2022}\natexlab{}.
\newblock \showarticletitle{Deep Learning Based Vulnerability Detection: Are We
  There Yet?}
\newblock \bibinfo{journal}{\emph{IEEE Transactions on Software Engineering}}
  \bibinfo{volume}{48}, \bibinfo{number}{9} (\bibinfo{year}{2022}),
  \bibinfo{pages}{3280--3296}.
\newblock
\urldef\tempurl%
\url{https://doi.org/10.1109/TSE.2021.3087402}
\showDOI{\tempurl}


\bibitem[Cheng et~al\mbox{.}(2021)]%
        {cheng_deepwukong_2021}
\bibfield{author}{\bibinfo{person}{Xiao Cheng}, \bibinfo{person}{Haoyu Wang},
  \bibinfo{person}{Jiayi Hua}, \bibinfo{person}{Guoai Xu}, {and}
  \bibinfo{person}{Yulei Sui}.} \bibinfo{year}{2021}\natexlab{}.
\newblock \showarticletitle{DeepWukong: Statically Detecting Software
  Vulnerabilities Using Deep Graph Neural Network}.
\newblock \bibinfo{journal}{\emph{ACM Trans. Softw. Eng. Methodol.}}
  \bibinfo{volume}{30}, \bibinfo{number}{3}, Article \bibinfo{articleno}{38}
  (\bibinfo{date}{apr} \bibinfo{year}{2021}), \bibinfo{numpages}{33}~pages.
\newblock
\showISSN{1049-331X}
\urldef\tempurl%
\url{https://doi.org/10.1145/3436877}
\showDOI{\tempurl}


\bibitem[Cheng et~al\mbox{.}(2022)]%
        {cheng_path-sensitive_2022}
\bibfield{author}{\bibinfo{person}{Xiao Cheng}, \bibinfo{person}{Guanqin
  Zhang}, \bibinfo{person}{Haoyu Wang}, {and} \bibinfo{person}{Yulei Sui}.}
  \bibinfo{year}{2022}\natexlab{}.
\newblock \showarticletitle{Path-Sensitive Code Embedding via Contrastive
  Learning for Software Vulnerability Detection}. In
  \bibinfo{booktitle}{\emph{Proceedings of the 31st ACM SIGSOFT International
  Symposium on Software Testing and Analysis}} (Virtual, South Korea)
  \emph{(\bibinfo{series}{ISSTA 2022})}. \bibinfo{publisher}{Association for
  Computing Machinery}, \bibinfo{address}{New York, NY, USA},
  \bibinfo{pages}{519–531}.
\newblock
\showISBNx{9781450393799}
\urldef\tempurl%
\url{https://doi.org/10.1145/3533767.3534371}
\showDOI{\tempurl}


\bibitem[Cranmer et~al\mbox{.}(2020)]%
        {cranmer_discovering_2020}
\bibfield{author}{\bibinfo{person}{Miles Cranmer}, \bibinfo{person}{Alvaro
  Sanchez~Gonzalez}, \bibinfo{person}{Peter Battaglia}, \bibinfo{person}{Rui
  Xu}, \bibinfo{person}{Kyle Cranmer}, \bibinfo{person}{David Spergel}, {and}
  \bibinfo{person}{Shirley Ho}.} \bibinfo{year}{2020}\natexlab{}.
\newblock \showarticletitle{Discovering Symbolic Models from Deep Learning with
  Inductive Biases}. In \bibinfo{booktitle}{\emph{Advances in Neural
  Information Processing Systems}},
  \bibfield{editor}{\bibinfo{person}{H.~Larochelle},
  \bibinfo{person}{M.~Ranzato}, \bibinfo{person}{R.~Hadsell},
  \bibinfo{person}{M.F. Balcan}, {and} \bibinfo{person}{H.~Lin}} (Eds.),
  Vol.~\bibinfo{volume}{33}. \bibinfo{publisher}{Curran Associates, Inc.},
  \bibinfo{pages}{17429--17442}.
\newblock
\urldef\tempurl%
\url{https://proceedings.neurips.cc/paper_files/paper/2020/file
  c9f2f917078bd2db12f23c3b413d9cba-Paper.pdf}
\showURL{%
\tempurl}


\bibitem[Cummins et~al\mbox{.}(2021)]%
        {cummins_programl_2021}
\bibfield{author}{\bibinfo{person}{Chris Cummins},
  \bibinfo{person}{Zacharias~V. Fisches}, \bibinfo{person}{Tal Ben-Nun},
  \bibinfo{person}{Torsten Hoefler}, \bibinfo{person}{Michael F~P O'Boyle},
  {and} \bibinfo{person}{Hugh Leather}.} \bibinfo{year}{2021}\natexlab{}.
\newblock \showarticletitle{ProGraML: A Graph-based Program Representation for
  Data Flow Analysis and Compiler Optimizations}. In
  \bibinfo{booktitle}{\emph{Proceedings of the 38th International Conference on
  Machine Learning}} \emph{(\bibinfo{series}{Proceedings of Machine Learning
  Research}, Vol.~\bibinfo{volume}{139})},
  \bibfield{editor}{\bibinfo{person}{Marina Meila} {and} \bibinfo{person}{Tong
  Zhang}} (Eds.). \bibinfo{publisher}{PMLR}, \bibinfo{pages}{2244--2253}.
\newblock
\urldef\tempurl%
\url{https://proceedings.mlr.press/v139/cummins21a.html}
\showURL{%
\tempurl}


\bibitem[Dietterich(1998)]%
        {dietterich_approximate_1998}
\bibfield{author}{\bibinfo{person}{Thomas~G. Dietterich}.}
  \bibinfo{year}{1998}\natexlab{}.
\newblock \showarticletitle{{Approximate Statistical Tests for Comparing
  Supervised Classification Learning Algorithms}}.
\newblock \bibinfo{journal}{\emph{Neural Computation}} \bibinfo{volume}{10},
  \bibinfo{number}{7} (\bibinfo{date}{10} \bibinfo{year}{1998}),
  \bibinfo{pages}{1895--1923}.
\newblock
\showISSN{0899-7667}
\urldef\tempurl%
\url{https://doi.org/10.1162/089976698300017197}
\showDOI{\tempurl}
\showeprint{https://direct.mit.edu/neco/article-pdf/10/7/1895/814002/089976698300017197.pdf}


\bibitem[Dinella et~al\mbox{.}(2020)]%
        {dinella_hoppity_2020}
\bibfield{author}{\bibinfo{person}{Elizabeth Dinella}, \bibinfo{person}{Hanjun
  Dai}, \bibinfo{person}{Ziyang Li}, \bibinfo{person}{Mayur Naik},
  \bibinfo{person}{Le Song}, {and} \bibinfo{person}{Ke Wang}.}
  \bibinfo{year}{2020}\natexlab{}.
\newblock \showarticletitle{HOPPITY: LEARNING GRAPH TRANSFORMATIONS TO DETECT
  AND FIX BUGS IN PROGRAMS}. In \bibinfo{booktitle}{\emph{International
  Conference on Learning Representations}}.
\newblock
\urldef\tempurl%
\url{https://openreview.net/forum?id=SJeqs6EFvB}
\showURL{%
\tempurl}


\bibitem[Ding et~al\mbox{.}(2022)]%
        {ding_velvet_2022}
\bibfield{author}{\bibinfo{person}{Y. Ding}, \bibinfo{person}{S. Suneja},
  \bibinfo{person}{Y. Zheng}, \bibinfo{person}{J. Laredo}, \bibinfo{person}{A.
  Morari}, \bibinfo{person}{G. Kaiser}, {and} \bibinfo{person}{B. Ray}.}
  \bibinfo{year}{2022}\natexlab{}.
\newblock \showarticletitle{VELVET: a noVel Ensemble Learning approach to
  automatically locate VulnErable sTatements}. In
  \bibinfo{booktitle}{\emph{2022 IEEE International Conference on Software
  Analysis, Evolution and Reengineering (SANER)}}. \bibinfo{publisher}{IEEE
  Computer Society}, \bibinfo{address}{Los Alamitos, CA, USA},
  \bibinfo{pages}{959--970}.
\newblock
\showISSN{1534-5351}
\urldef\tempurl%
\url{https://doi.org/10.1109/SANER53432.2022.00114}
\showDOI{\tempurl}


\bibitem[Du et~al\mbox{.}(2019)]%
        {du_leopard_2020}
\bibfield{author}{\bibinfo{person}{Xiaoning Du}, \bibinfo{person}{Bihuan Chen},
  \bibinfo{person}{Yuekang Li}, \bibinfo{person}{Jianmin Guo},
  \bibinfo{person}{Yaqin Zhou}, \bibinfo{person}{Yang Liu}, {and}
  \bibinfo{person}{Yu Jiang}.} \bibinfo{year}{2019}\natexlab{}.
\newblock \showarticletitle{Leopard: Identifying Vulnerable Code for
  Vulnerability Assessment through Program Metrics}. In
  \bibinfo{booktitle}{\emph{Proceedings of the 41st International Conference on
  Software Engineering}} (Montreal, Quebec, Canada)
  \emph{(\bibinfo{series}{ICSE '19})}. \bibinfo{publisher}{IEEE Press},
  \bibinfo{pages}{60–71}.
\newblock
\urldef\tempurl%
\url{https://doi.org/10.1109/ICSE.2019.00024}
\showDOI{\tempurl}


\bibitem[Fan et~al\mbox{.}(2020)]%
        {fan_cc_2020}
\bibfield{author}{\bibinfo{person}{Jiahao Fan}, \bibinfo{person}{Yi Li},
  \bibinfo{person}{Shaohua Wang}, {and} \bibinfo{person}{Tien~N. Nguyen}.}
  \bibinfo{year}{2020}\natexlab{}.
\newblock \showarticletitle{A C/C++ Code Vulnerability Dataset with Code
  Changes and CVE Summaries}. In \bibinfo{booktitle}{\emph{Proceedings of the
  17th International Conference on Mining Software Repositories}} (Seoul,
  Republic of Korea) \emph{(\bibinfo{series}{MSR '20})}.
  \bibinfo{publisher}{Association for Computing Machinery},
  \bibinfo{address}{New York, NY, USA}, \bibinfo{pages}{508–512}.
\newblock
\showISBNx{9781450375177}
\urldef\tempurl%
\url{https://doi.org/10.1145/3379597.3387501}
\showDOI{\tempurl}


\bibitem[Feng et~al\mbox{.}(2020)]%
        {feng_codebert_2020}
\bibfield{author}{\bibinfo{person}{Zhangyin Feng}, \bibinfo{person}{Daya Guo},
  \bibinfo{person}{Duyu Tang}, \bibinfo{person}{Nan Duan},
  \bibinfo{person}{Xiaocheng Feng}, \bibinfo{person}{Ming Gong},
  \bibinfo{person}{Linjun Shou}, \bibinfo{person}{Bing Qin},
  \bibinfo{person}{Ting Liu}, \bibinfo{person}{Daxin Jiang}, {and}
  \bibinfo{person}{Ming Zhou}.} \bibinfo{year}{2020}\natexlab{}.
\newblock \showarticletitle{{C}ode{BERT}: A Pre-Trained Model for Programming
  and Natural Languages}. In \bibinfo{booktitle}{\emph{Findings of the
  Association for Computational Linguistics: EMNLP 2020}}.
  \bibinfo{publisher}{Association for Computational Linguistics},
  \bibinfo{address}{Online}, \bibinfo{pages}{1536--1547}.
\newblock
\urldef\tempurl%
\url{https://doi.org/10.18653/v1/2020.findings-emnlp.139}
\showDOI{\tempurl}


\bibitem[Fu and Tantithamthavorn(2022)]%
        {fu_linevul_2022}
\bibfield{author}{\bibinfo{person}{Michael Fu} {and} \bibinfo{person}{Chakkrit
  Tantithamthavorn}.} \bibinfo{year}{2022}\natexlab{}.
\newblock \showarticletitle{LineVul: A Transformer-Based Line-Level
  Vulnerability Prediction}. In \bibinfo{booktitle}{\emph{Proceedings of the
  19th International Conference on Mining Software Repositories}} (Pittsburgh,
  Pennsylvania) \emph{(\bibinfo{series}{MSR '22})}.
  \bibinfo{publisher}{Association for Computing Machinery},
  \bibinfo{address}{New York, NY, USA}, \bibinfo{pages}{608–620}.
\newblock
\showISBNx{9781450393034}
\urldef\tempurl%
\url{https://doi.org/10.1145/3524842.3528452}
\showDOI{\tempurl}


\bibitem[Gilmer et~al\mbox{.}(2017)]%
        {gilmerNeuralMessagePassing2017a}
\bibfield{author}{\bibinfo{person}{Justin Gilmer}, \bibinfo{person}{Samuel~S.
  Schoenholz}, \bibinfo{person}{Patrick~F. Riley}, \bibinfo{person}{Oriol
  Vinyals}, {and} \bibinfo{person}{George~E. Dahl}.}
  \bibinfo{year}{2017}\natexlab{}.
\newblock \bibinfo{title}{Neural Message Passing for Quantum Chemistry}.
\newblock
\newblock
\showeprint[arxiv]{1704.01212}~[cs.LG]


\bibitem[Guo et~al\mbox{.}(2022)]%
        {guo_unixcoder_2022}
\bibfield{author}{\bibinfo{person}{Daya Guo}, \bibinfo{person}{Shuai Lu},
  \bibinfo{person}{Nan Duan}, \bibinfo{person}{Yanlin Wang},
  \bibinfo{person}{Ming Zhou}, {and} \bibinfo{person}{Jian Yin}.}
  \bibinfo{year}{2022}\natexlab{}.
\newblock \bibinfo{title}{UniXcoder: Unified Cross-Modal Pre-training for Code
  Representation}.
\newblock
\newblock
\showeprint[arxiv]{2203.03850}~[cs.CL]


\bibitem[Hanif and Maffeis(2022)]%
        {hanif_vulberta_2022}
\bibfield{author}{\bibinfo{person}{Hazim Hanif} {and} \bibinfo{person}{Sergio
  Maffeis}.} \bibinfo{year}{2022}\natexlab{}.
\newblock \showarticletitle{VulBERTa: Simplified Source Code Pre-Training for
  Vulnerability Detection}. In \bibinfo{booktitle}{\emph{2022 International
  Joint Conference on Neural Networks (IJCNN)}}. \bibinfo{pages}{1--8}.
\newblock
\urldef\tempurl%
\url{https://doi.org/10.1109/IJCNN55064.2022.9892280}
\showDOI{\tempurl}


\bibitem[Hin et~al\mbox{.}(2022)]%
        {hin_linevd_2022}
\bibfield{author}{\bibinfo{person}{David Hin}, \bibinfo{person}{Andrey Kan},
  \bibinfo{person}{Huaming Chen}, {and} \bibinfo{person}{M.~Ali Babar}.}
  \bibinfo{year}{2022}\natexlab{}.
\newblock \showarticletitle{LineVD: Statement-Level Vulnerability Detection
  Using Graph Neural Networks}. In \bibinfo{booktitle}{\emph{Proceedings of the
  19th International Conference on Mining Software Repositories}} (Pittsburgh,
  Pennsylvania) \emph{(\bibinfo{series}{MSR '22})}.
  \bibinfo{publisher}{Association for Computing Machinery},
  \bibinfo{address}{New York, NY, USA}, \bibinfo{pages}{596–607}.
\newblock
\showISBNx{9781450393034}
\urldef\tempurl%
\url{https://doi.org/10.1145/3524842.3527949}
\showDOI{\tempurl}


\bibitem[Horwitz({[n.\,d.]})]%
        {Horwitz:dataflow}
\bibfield{author}{\bibinfo{person}{Susan Horwitz}.}
  \bibinfo{year}{[n.\,d.]}\natexlab{}.
\newblock \bibinfo{booktitle}{\emph{Dataflow Analysis}}.
\newblock
\urldef\tempurl%
\url{https://pages.cs.wisc.edu/~horwitz/CS704-NOTES/2.DATAFLOW.html}
\showURL{%
\tempurl}
\newblock
\shownote{CS704 Lecture Notes [Accessed 19-09-2023]}.


\bibitem[Howard et~al\mbox{.}(2017)]%
        {howard_mobilenets_2017}
\bibfield{author}{\bibinfo{person}{Andrew~G. Howard}, \bibinfo{person}{Menglong
  Zhu}, \bibinfo{person}{Bo Chen}, \bibinfo{person}{Dmitry Kalenichenko},
  \bibinfo{person}{Weijun Wang}, \bibinfo{person}{Tobias Weyand},
  \bibinfo{person}{Marco Andreetto}, {and} \bibinfo{person}{Hartwig Adam}.}
  \bibinfo{year}{2017}\natexlab{}.
\newblock \bibinfo{title}{MobileNets: Efficient Convolutional Neural Networks
  for Mobile Vision Applications}.
\newblock
\newblock
\showeprint[arxiv]{1704.04861}~[cs.CV]


\bibitem[Japkowicz(2000)]%
        {japkowicz2000class}
\bibfield{author}{\bibinfo{person}{Nathalie Japkowicz}.}
  \bibinfo{year}{2000}\natexlab{}.
\newblock \showarticletitle{The class imbalance problem: Significance and
  strategies}. In \bibinfo{booktitle}{\emph{Proc. of the Int’l Conf. on
  artificial intelligence}}, Vol.~\bibinfo{volume}{56}.
  \bibinfo{pages}{111--117}.
\newblock


\bibitem[Jeon et~al\mbox{.}(2019)]%
        {jeon_machine-learning_2019}
\bibfield{author}{\bibinfo{person}{Minseok Jeon}, \bibinfo{person}{Sehun
  Jeong}, \bibinfo{person}{Sungdeok Cha}, {and} \bibinfo{person}{Hakjoo Oh}.}
  \bibinfo{year}{2019}\natexlab{}.
\newblock \showarticletitle{A Machine-Learning Algorithm with Disjunctive Model
  for Data-Driven Program Analysis}.
\newblock \bibinfo{journal}{\emph{ACM Trans. Program. Lang. Syst.}}
  \bibinfo{volume}{41}, \bibinfo{number}{2}, Article \bibinfo{articleno}{13}
  (\bibinfo{date}{June} \bibinfo{year}{2019}), \bibinfo{numpages}{41}~pages.
\newblock
\showISSN{0164-0925}
\urldef\tempurl%
\url{https://doi.org/10.1145/3293607}
\showDOI{\tempurl}


\bibitem[Kildall(1973)]%
        {kildall_unified_1973}
\bibfield{author}{\bibinfo{person}{Gary~A. Kildall}.}
  \bibinfo{year}{1973}\natexlab{}.
\newblock \showarticletitle{A Unified Approach to Global Program Optimization}.
  In \bibinfo{booktitle}{\emph{Proceedings of the 1st Annual ACM SIGACT-SIGPLAN
  Symposium on Principles of Programming Languages}} (Boston, Massachusetts)
  \emph{(\bibinfo{series}{POPL '73})}. \bibinfo{publisher}{Association for
  Computing Machinery}, \bibinfo{address}{New York, NY, USA},
  \bibinfo{pages}{194–206}.
\newblock
\showISBNx{9781450373494}
\urldef\tempurl%
\url{https://doi.org/10.1145/512927.512945}
\showDOI{\tempurl}


\bibitem[Le and Mikolov(2014)]%
        {le_distributed_2014}
\bibfield{author}{\bibinfo{person}{Quoc~V. Le} {and} \bibinfo{person}{Tomas
  Mikolov}.} \bibinfo{year}{2014}\natexlab{}.
\newblock \bibinfo{title}{Distributed Representations of Sentences and
  Documents}.
\newblock
\newblock
\showeprint[arxiv]{1405.4053}~[cs.CL]


\bibitem[Li et~al\mbox{.}(2017)]%
        {li_gated_2016}
\bibfield{author}{\bibinfo{person}{Yujia Li}, \bibinfo{person}{Daniel Tarlow},
  \bibinfo{person}{Marc Brockschmidt}, {and} \bibinfo{person}{Richard Zemel}.}
  \bibinfo{year}{2017}\natexlab{}.
\newblock \bibinfo{title}{Gated Graph Sequence Neural Networks}.
\newblock
\newblock
\showeprint[arxiv]{1511.05493}~[cs.LG]


\bibitem[Li et~al\mbox{.}(2021)]%
        {li_vulnerability_2021}
\bibfield{author}{\bibinfo{person}{Yi Li}, \bibinfo{person}{Shaohua Wang},
  {and} \bibinfo{person}{Tien~N. Nguyen}.} \bibinfo{year}{2021}\natexlab{}.
\newblock \showarticletitle{Vulnerability detection with fine-grained
  interpretations}.
\newblock  (\bibinfo{year}{2021}), \bibinfo{pages}{292--303}.
\newblock
\showISSN{9781450385626}
\urldef\tempurl%
\url{https://doi.org/10.1145/3468264.3468597}
\showDOI{\tempurl}


\bibitem[Li et~al\mbox{.}(2022)]%
        {li_sysevr_2021}
\bibfield{author}{\bibinfo{person}{Zhen Li}, \bibinfo{person}{Deqing Zou},
  \bibinfo{person}{Shouhuai Xu}, \bibinfo{person}{Hai Jin},
  \bibinfo{person}{Yawei Zhu}, {and} \bibinfo{person}{Zhaoxuan Chen}.}
  \bibinfo{year}{2022}\natexlab{}.
\newblock \showarticletitle{SySeVR: A Framework for Using Deep Learning to
  Detect Software Vulnerabilities}.
\newblock \bibinfo{journal}{\emph{IEEE Transactions on Dependable and Secure
  Computing}} \bibinfo{volume}{19}, \bibinfo{number}{4} (\bibinfo{year}{2022}),
  \bibinfo{pages}{2244--2258}.
\newblock
\urldef\tempurl%
\url{https://doi.org/10.1109/TDSC.2021.3051525}
\showDOI{\tempurl}


\bibitem[Li et~al\mbox{.}(2018)]%
        {li_vuldeepecker_2018}
\bibfield{author}{\bibinfo{person}{Zhen Li}, \bibinfo{person}{Deqing Zou},
  \bibinfo{person}{Shouhuai Xu}, \bibinfo{person}{Xinyu Ou},
  \bibinfo{person}{Hai Jin}, \bibinfo{person}{Sujuan Wang},
  \bibinfo{person}{Zhijun Deng}, {and} \bibinfo{person}{Yuyi Zhong}.}
  \bibinfo{year}{2018}\natexlab{}.
\newblock \showarticletitle{{VulDeePecker}: {A} {Deep} {Learning}-{Based}
  {System} for {Vulnerability} {Detection}}.
\newblock  \bibinfo{number}{February} (\bibinfo{year}{2018}).
\newblock
\urldef\tempurl%
\url{https://doi.org/10.14722/ndss.2018.23158}
\showDOI{\tempurl}


\bibitem[Lu et~al\mbox{.}(2021)]%
        {lu_codexglue_2021}
\bibfield{author}{\bibinfo{person}{Shuai Lu}, \bibinfo{person}{Daya Guo},
  \bibinfo{person}{Shuo Ren}, \bibinfo{person}{Junjie Huang},
  \bibinfo{person}{Alexey Svyatkovskiy}, \bibinfo{person}{Ambrosio Blanco},
  \bibinfo{person}{Colin~B. Clement}, \bibinfo{person}{Dawn Drain},
  \bibinfo{person}{Daxin Jiang}, \bibinfo{person}{Duyu Tang},
  \bibinfo{person}{Ge Li}, \bibinfo{person}{Lidong Zhou},
  \bibinfo{person}{Linjun Shou}, \bibinfo{person}{Long Zhou},
  \bibinfo{person}{Michele Tufano}, \bibinfo{person}{Ming Gong},
  \bibinfo{person}{Ming Zhou}, \bibinfo{person}{Nan Duan},
  \bibinfo{person}{Neel Sundaresan}, \bibinfo{person}{Shao~Kun Deng},
  \bibinfo{person}{Shengyu Fu}, {and} \bibinfo{person}{Shujie Liu}.}
  \bibinfo{year}{2021}\natexlab{}.
\newblock \showarticletitle{CodeXGLUE: {A} Machine Learning Benchmark Dataset
  for Code Understanding and Generation}.
\newblock \bibinfo{journal}{\emph{CoRR}}  \bibinfo{volume}{abs/2102.04664}
  (\bibinfo{year}{2021}).
\newblock


\bibitem[Mikolov et~al\mbox{.}(2013)]%
        {mikolov_efficient_2013}
\bibfield{author}{\bibinfo{person}{Tomas Mikolov}, \bibinfo{person}{Kai Chen},
  \bibinfo{person}{Greg Corrado}, {and} \bibinfo{person}{Jeffrey Dean}.}
  \bibinfo{year}{2013}\natexlab{}.
\newblock \bibinfo{title}{Efficient Estimation of Word Representations in
  Vector Space}.
\newblock
\newblock
\showeprint[arxiv]{1301.3781}~[cs.CL]


\bibitem[Moshtari et~al\mbox{.}(2022)]%
        {moshtari_grounded_2022}
\bibfield{author}{\bibinfo{person}{Sara Moshtari}, \bibinfo{person}{Ahmet
  Okutan}, {and} \bibinfo{person}{Mehdi Mirakhorli}.}
  \bibinfo{year}{2022}\natexlab{}.
\newblock \showarticletitle{A Grounded Theory Based Approach to Characterize
  Software Attack Surfaces}. In \bibinfo{booktitle}{\emph{Proceedings of the
  44th International Conference on Software Engineering}} (Pittsburgh,
  Pennsylvania) \emph{(\bibinfo{series}{ICSE '22})}.
  \bibinfo{publisher}{Association for Computing Machinery},
  \bibinfo{address}{New York, NY, USA}, \bibinfo{pages}{13–24}.
\newblock
\showISBNx{9781450392211}
\urldef\tempurl%
\url{https://doi.org/10.1145/3510003.3510210}
\showDOI{\tempurl}


\bibitem[Nguyen et~al\mbox{.}(2022)]%
        {nguyen_regvd_2022}
\bibfield{author}{\bibinfo{person}{Van-Anh Nguyen}, \bibinfo{person}{Dai~Quoc
  Nguyen}, \bibinfo{person}{Van Nguyen}, \bibinfo{person}{Trung Le},
  \bibinfo{person}{Quan~Hung Tran}, {and} \bibinfo{person}{Dinh Phung}.}
  \bibinfo{year}{2022}\natexlab{}.
\newblock \showarticletitle{Re{GVD}: Revisiting Graph Neural Networks for
  Vulnerability Detection}. In \bibinfo{booktitle}{\emph{Deep Learning for Code
  Workshop}}.
\newblock
\urldef\tempurl%
\url{https://openreview.net/forum?id=BU5eniuWkbq}
\showURL{%
\tempurl}


\bibitem[Pennington et~al\mbox{.}(2014)]%
        {pennington_glove_2014}
\bibfield{author}{\bibinfo{person}{Jeffrey Pennington},
  \bibinfo{person}{Richard Socher}, {and} \bibinfo{person}{Christopher
  Manning}.} \bibinfo{year}{2014}\natexlab{}.
\newblock \showarticletitle{{G}lo{V}e: Global Vectors for Word Representation}.
  In \bibinfo{booktitle}{\emph{Proceedings of the 2014 Conference on Empirical
  Methods in Natural Language Processing ({EMNLP})}}.
  \bibinfo{publisher}{Association for Computational Linguistics},
  \bibinfo{address}{Doha, Qatar}, \bibinfo{pages}{1532--1543}.
\newblock
\urldef\tempurl%
\url{https://doi.org/10.3115/v1/D14-1162}
\showDOI{\tempurl}


\bibitem[Rasley et~al\mbox{.}(2020)]%
        {deepspeed}
\bibfield{author}{\bibinfo{person}{Jeff Rasley}, \bibinfo{person}{Samyam
  Rajbhandari}, \bibinfo{person}{Olatunji Ruwase}, {and}
  \bibinfo{person}{Yuxiong He}.} \bibinfo{year}{2020}\natexlab{}.
\newblock \showarticletitle{DeepSpeed: System Optimizations Enable Training
  Deep Learning Models with Over 100 Billion Parameters}. In
  \bibinfo{booktitle}{\emph{Proceedings of the 26th ACM SIGKDD International
  Conference on Knowledge Discovery \& Data Mining}} (Virtual Event, CA, USA)
  \emph{(\bibinfo{series}{KDD '20})}. \bibinfo{publisher}{Association for
  Computing Machinery}, \bibinfo{address}{New York, NY, USA},
  \bibinfo{pages}{3505–3506}.
\newblock
\showISBNx{9781450379984}
\urldef\tempurl%
\url{https://doi.org/10.1145/3394486.3406703}
\showDOI{\tempurl}


\bibitem[Reps et~al\mbox{.}(1995)]%
        {repsPreciseInterproceduralDataflow1995}
\bibfield{author}{\bibinfo{person}{Thomas Reps}, \bibinfo{person}{Susan
  Horwitz}, {and} \bibinfo{person}{Mooly Sagiv}.}
  \bibinfo{year}{1995}\natexlab{}.
\newblock \showarticletitle{Precise Interprocedural Dataflow Analysis via Graph
  Reachability}. In \bibinfo{booktitle}{\emph{Proceedings of the 22nd {{ACM
  SIGPLAN-SIGACT}} Symposium on {{Principles}} of Programming Languages}}
  \emph{(\bibinfo{series}{{{POPL}} '95})}. \bibinfo{publisher}{{Association for
  Computing Machinery}}, \bibinfo{address}{{New York, NY, USA}},
  \bibinfo{pages}{49--61}.
\newblock
\showISBNx{978-0-89791-692-9}
\urldef\tempurl%
\url{https://doi.org/10.1145/199448.199462}
\showDOI{\tempurl}


\bibitem[Russell et~al\mbox{.}(2018)]%
        {russell_automated_2019}
\bibfield{author}{\bibinfo{person}{Rebecca Russell}, \bibinfo{person}{Louis
  Kim}, \bibinfo{person}{Lei Hamilton}, \bibinfo{person}{Tomo Lazovich},
  \bibinfo{person}{Jacob Harer}, \bibinfo{person}{Onur Ozdemir},
  \bibinfo{person}{Paul Ellingwood}, {and} \bibinfo{person}{Marc McConley}.}
  \bibinfo{year}{2018}\natexlab{}.
\newblock \showarticletitle{Automated Vulnerability Detection in Source Code
  Using Deep Representation Learning}. In \bibinfo{booktitle}{\emph{2018 17th
  IEEE International Conference on Machine Learning and Applications (ICMLA)}}.
  \bibinfo{pages}{757--762}.
\newblock
\urldef\tempurl%
\url{https://doi.org/10.1109/ICMLA.2018.00120}
\showDOI{\tempurl}


\bibitem[Seabold and Perktold(2010)]%
        {seabold2010statsmodels}
\bibfield{author}{\bibinfo{person}{Skipper Seabold} {and}
  \bibinfo{person}{Josef Perktold}.} \bibinfo{year}{2010}\natexlab{}.
\newblock \showarticletitle{statsmodels: Econometric and statistical modeling
  with python}. In \bibinfo{booktitle}{\emph{9th Python in Science
  Conference}}.
\newblock


\bibitem[Steenhoek et~al\mbox{.}(2023)]%
        {empiricalstudy}
\bibfield{author}{\bibinfo{person}{Benjamin Steenhoek},
  \bibinfo{person}{Md~Mahbubur Rahman}, \bibinfo{person}{Richard Jiles}, {and}
  \bibinfo{person}{Wei Le}.} \bibinfo{year}{2023}\natexlab{}.
\newblock \showarticletitle{An Empirical Study of Deep Learning Models for
  Vulnerability Detection}. In \bibinfo{booktitle}{\emph{2023 IEEE/ACM 45th
  International Conference on Software Engineering (ICSE)}}.
  \bibinfo{pages}{2237--2248}.
\newblock
\urldef\tempurl%
\url{https://doi.org/10.1109/ICSE48619.2023.00188}
\showDOI{\tempurl}


\bibitem[Tan and Le(2019)]%
        {tan_efficientnet_2020}
\bibfield{author}{\bibinfo{person}{Mingxing Tan} {and} \bibinfo{person}{Quoc
  Le}.} \bibinfo{year}{2019}\natexlab{}.
\newblock \showarticletitle{{E}fficient{N}et: Rethinking Model Scaling for
  Convolutional Neural Networks}. In \bibinfo{booktitle}{\emph{Proceedings of
  the 36th International Conference on Machine Learning}}
  \emph{(\bibinfo{series}{Proceedings of Machine Learning Research},
  Vol.~\bibinfo{volume}{97})}, \bibfield{editor}{\bibinfo{person}{Kamalika
  Chaudhuri} {and} \bibinfo{person}{Ruslan Salakhutdinov}} (Eds.).
  \bibinfo{publisher}{PMLR}, \bibinfo{pages}{6105--6114}.
\newblock
\urldef\tempurl%
\url{https://proceedings.mlr.press/v97/tan19a.html}
\showURL{%
\tempurl}


\bibitem[VenkataKeerthy et~al\mbox{.}(2020)]%
        {venkatakeerthy_ir2vec_2020}
\bibfield{author}{\bibinfo{person}{S. VenkataKeerthy}, \bibinfo{person}{Rohit
  Aggarwal}, \bibinfo{person}{Shalini Jain}, \bibinfo{person}{Maunendra~Sankar
  Desarkar}, \bibinfo{person}{Ramakrishna Upadrasta}, {and}
  \bibinfo{person}{Y.~N. Srikant}.} \bibinfo{year}{2020}\natexlab{}.
\newblock \showarticletitle{IR2VEC: LLVM IR Based Scalable Program Embeddings}.
\newblock \bibinfo{journal}{\emph{ACM Trans. Archit. Code Optim.}}
  \bibinfo{volume}{17}, \bibinfo{number}{4}, Article \bibinfo{articleno}{32}
  (\bibinfo{date}{dec} \bibinfo{year}{2020}), \bibinfo{numpages}{27}~pages.
\newblock
\showISSN{1544-3566}
\urldef\tempurl%
\url{https://doi.org/10.1145/3418463}
\showDOI{\tempurl}


\bibitem[Wang et~al\mbox{.}(2020)]%
        {wang_learning_2020}
\bibfield{author}{\bibinfo{person}{Yu Wang}, \bibinfo{person}{Fengjuan Gao},
  \bibinfo{person}{Linzhang Wang}, {and} \bibinfo{person}{Ke Wang}.}
  \bibinfo{year}{2020}\natexlab{}.
\newblock \showarticletitle{Learning a {Static} {Bug} {Finder} from {Data}}.
\newblock \bibinfo{journal}{\emph{arXiv:1907.05579 [cs]}}
  (\bibinfo{date}{March} \bibinfo{year}{2020}).
\newblock
\urldef\tempurl%
\url{http://arxiv.org/abs/1907.05579}
\showURL{%
\tempurl}
\newblock
\shownote{arXiv: 1907.05579}.


\bibitem[Wikipedia(2021)]%
        {wikipedia_databreaches_2021}
\bibfield{author}{\bibinfo{person}{Wikipedia}.}
  \bibinfo{year}{2021}\natexlab{}.
\newblock \bibinfo{title}{List of data breaches}.
\newblock
  \bibinfo{howpublished}{\url{https://web.archive.org/web/20211011144237/https://en.wikipedia.org/wiki/List_of_data_breaches}}.
\newblock
\newblock
\shownote{Accessed October 29 2021}.


\bibitem[Xu et~al\mbox{.}(2019)]%
        {xu_how_2019}
\bibfield{author}{\bibinfo{person}{Keyulu Xu}, \bibinfo{person}{Weihua Hu},
  \bibinfo{person}{Jure Leskovec}, {and} \bibinfo{person}{Stefanie Jegelka}.}
  \bibinfo{year}{2019}\natexlab{}.
\newblock \showarticletitle{How Powerful are Graph Neural Networks?}. In
  \bibinfo{booktitle}{\emph{International Conference on Learning
  Representations}}.
\newblock
\urldef\tempurl%
\url{https://openreview.net/forum?id=ryGs6iA5Km}
\showURL{%
\tempurl}


\bibitem[Yamaguchi et~al\mbox{.}(2014)]%
        {yamaguchi_modeling_2014}
\bibfield{author}{\bibinfo{person}{Fabian Yamaguchi}, \bibinfo{person}{Nico
  Golde}, \bibinfo{person}{Daniel Arp}, {and} \bibinfo{person}{Konrad Rieck}.}
  \bibinfo{year}{2014}\natexlab{}.
\newblock \showarticletitle{Modeling and discovering vulnerabilities with code
  property graphs}.
\newblock \bibinfo{journal}{\emph{Proceedings - IEEE Symposium on Security and
  Privacy}} (\bibinfo{year}{2014}), \bibinfo{pages}{590--604}.
\newblock
\showISSN{9781479946860}
\urldef\tempurl%
\url{https://doi.org/10.1109/SP.2014.44}
\showDOI{\tempurl}


\bibitem[Zheng et~al\mbox{.}(2021)]%
        {zheng_d2a_2021}
\bibfield{author}{\bibinfo{person}{Yunhui Zheng}, \bibinfo{person}{Saurabh
  Pujar}, \bibinfo{person}{Burn Lewis}, \bibinfo{person}{Luca Buratti},
  \bibinfo{person}{Edward Epstein}, \bibinfo{person}{Bo Yang},
  \bibinfo{person}{Jim Laredo}, \bibinfo{person}{Alessandro Morari}, {and}
  \bibinfo{person}{Zhong Su}.} \bibinfo{year}{2021}\natexlab{}.
\newblock \showarticletitle{{D2A}: {A} {Dataset} {Built} for {AI}-{Based}
  {Vulnerability} {Detection} {Methods} {Using} {Differential} {Analysis}}.
\newblock  (\bibinfo{year}{2021}), \bibinfo{pages}{111--120}.
\newblock
\urldef\tempurl%
\url{https://doi.org/10.1109/icse-seip52600.2021.00020}
\showDOI{\tempurl}


\bibitem[Zhou et~al\mbox{.}(2019)]%
        {zhou_devign_2019}
\bibfield{author}{\bibinfo{person}{Yaqin Zhou}, \bibinfo{person}{Shangqing
  Liu}, \bibinfo{person}{Jingkai Siow}, \bibinfo{person}{Xiaoning Du}, {and}
  \bibinfo{person}{Yang Liu}.} \bibinfo{year}{2019}\natexlab{}.
\newblock \showarticletitle{Devign: {Effective} vulnerability identification by
  learning comprehensive program semantics via graph neural networks}.
\newblock \bibinfo{journal}{\emph{Advances in Neural Information Processing
  Systems}}  \bibinfo{volume}{32} (\bibinfo{year}{2019}),
  \bibinfo{pages}{1--11}.
\newblock
\urldef\tempurl%
\url{https://doi.org/10.5555/3454287.3455202}
\showDOI{\tempurl}


\end{thebibliography}


\addtocounter{TotPages}{-2} 

\clearpage

\appendix


\twocolumn[\centering
\textbf{\huge Supplemental Materials: Dataflow Analysis-Inspired Deep Learning for Efficient Vulnerability Detection
}
\vspace{0.25in}]
\begin{appendices}



\section{Baseline reproductions}
\label{sec:training-appendix}





\begin{itemize}
    \item We could not reproduce VulDeePecker, SySeVR, Draper, or IVDetect, so we repeated the performances reported in  \citet{li_vulnerability_2021}. Their measurements may vary slightly from our reproduction.
    \item We confirmed with \citet{chakraborty_deep_2022} that our results fixed a data leakage bug in the original implementation, so the results we report may differ from the original paper.
    \item \citet{zhou_devign_2019} did not release their model code, so we reproduced Devign from the third-party implementation released by \citet{chakraborty_deep_2022} (\url{https://github.com/saikat107/Devign}).
    \item \citet{hin_linevd_2022} did not report function-level metrics for LineVD, and we could not reproduce their statement-level performance from their model code, so we did not compare with their approach.
    \item \citet{cheng_path-sensitive_2022} did not release their model code and evaluated a different dataset than ours, so we did not compare with their approach.
\end{itemize}

\section{Programs that are removed}\label{sec:appendix-joern-failures}

We excluded a total of 1,564 programs (0.8\%) of the Big-Vul dataset, following the implementation of LineVD\footnote{\url{https://github.com/davidhin/linevd}}. Specifically, the following programs are removed:
\begin{itemize}
    \item Incomplete functions, i.e., ending with `\texttt{);}', or not ending in `\texttt{\}}' or no `\texttt{;}'. Joern can fail to parse such functions.
    \item Programs where no lines were added or removed, but the function was labeled vulnerable.
    \item Vulnerable programs where more than $70\%$ of the lines are modified from the vulnerable to the fixed version, indicating that a substantial change has been done and the vulnerability may be changed.
    \item Programs that are fewer than 5 lines long.
\end{itemize}

\newpage

\section{Additional effectiveness results}
\label{sec:appendix-100-1}

\renewcommand{\mytablespace}{\cmidrule{2-5}}

Of the transformer models, we evaluated CodeBERT and LineVul in our experiment for Section \ref{sec:effectiveness}.
Table \ref{fig:extra-effectiveness} reports the performances of the other models.

\begin{table}[htbp]
\centering
\caption{Initial trial run of performance on 100\% of the {\dataset} dataset.
}\label{fig:extra-effectiveness}
\begin{tabular}{llccc}
\toprule
Model & Model type       & F1 & Precision & Recall \\
\midrule
Devign  & GNN & \makecell[c]{29.33\\(6.58)} & \makecell[c]{32.83\\(5.55)}  & \makecell[c]{26.59\\(7.25)} \\
\mytablespace
ReVeal  & GNN      & \makecell[c]{33.60\\ (0.69)}  & \makecell[c]{33.08\\ (3.49)}  & \makecell[c]{34.67\\ (3.80)}  \\
\mytablespace
ReGVD   & GNN       & \makecell[c]{24.49\\ (3.16)} & \makecell[c]{63.76\\ (3.54)}  & \makecell[c]{15.26\\ (2.71)} \\
\mytablespace
CodeBERT & Transformer    & \makecell[c]{22.68\\ (8.12)} & \makecell[c]{67.79\\ (4.90)}   & \makecell[c]{19.11\\ (3.39)} \\
\mytablespace
LineVul  & Transformer     & \makecell[c]{\textbf{91.58}\\ (0.49)} & \makecell[c]{\textbf{95.99}\\ (0.85)}  & \makecell[c]{\textbf{87.55}\\ (0.49)} \\
\mytablespace
VulBERTaMLP~\citep{hanif_vulberta_2022}  & Transformer & \makecell[c]{1.75\\ (3.03)}  & \makecell[c]{19.33\\ (33.49)} & \makecell[c]{0.92\\ (1.59)}  \\
\mytablespace
VulBERTaCNN~\citep{hanif_vulberta_2022}  & Transformer & \makecell[c]{10.59\\ (0.00)}    & \makecell[c]{5.59\\ (0.00)}      & \makecell[c]{100.00\\ (0.00)}      \\
\mytablespace
PLBART~\citep{ahmad_unified_2021}  & Transformer      & \makecell[c]{25.35\\ (3.74)} & \makecell[c]{61.84\\ (6.54)}  & \makecell[c]{16.18\\ (3.52)} \\
\bottomrule
\end{tabular}
\end{table}


\clearpage


\section{Training times of all models}
\label{sec:all-training-times}

Table \ref{fig:all-training-time} lists the training times of the lower-performing models which we did not report in Section \ref{sec:efficiency}.
These training runs were not performed in the same environment as those in Section \ref{sec:efficiency}, but they provide an approximate measurement of the training time. We were not able to reproduce IVDetect, so we do not report its training time.

\begin{table}[H]
    \centering
    \caption{Approximate training times of all models.
    }
    \label{fig:all-training-time}
    \begin{tabular}{lr}
    \toprule
    Model     & Training time \\
    \midrule
    Devign    & 2h58m \\
    ReVeal    & 13h20m \\
    ReGVD     & 5h33m \\
    CodeBERT  & 7h33m \\
    LineVul   & 10h19m \\
    {\tool{}+LineVul} & 10h40m \\
    UniXcoder & 11h16m \\
    {\tool{}+UniXcoder} & 13h22m \\
    CodeT5 & 27h40m \\
    {\tool{}+CodeT5} & 27h57m \\
    {\tool{}}   & \textbf{9m} \\
    \bottomrule
    \end{tabular}
\end{table}

\section{Model sizes}
\label{sec:model-sizes}

Table \ref{fig:parameters} lists the size of each model.

\begin{table}[H]
\centering
\caption{{\tool} was smallest in terms of parameter count.}\label{fig:parameters}
\begin{tabular}{lr}
\toprule
Model & \# parameters \\
\midrule
IVDetect        & 924,165       \\
Devign          & 1,148,553     \\
ReVeal          & 560,291       \\
ReGVD           & 124,794,500   \\
CodeBERT        & 124,646,401   \\
LineVul         & 125,238,531   \\
{\tool{}+LineVul}       & 125,679,236         \\
UniXcoder         & 126,522,627   \\
{\tool{}+UniXcoder}       & 126,963,332         \\
CodeT5         & 222,883,586   \\
{\tool{}+CodeT5}         & 223,128,195   \\
{\tool}         & \textbf{375,938}       \\
\bottomrule
\end{tabular}

\end{table}

\newpage

\section{Additional cross-project evaluation results}
\label{sec:appendix-100-2}


We only evaluated LineVul in our experiment for Section \ref{sec:generalization} because its absolute performance was substantially better than the other baseline models in our initial trial. Table \ref{fig:100-percent-cross-project} reports the results of our initial trial.

\begin{table}[H]
\centering
\caption{Initial trial run of cross-project evaluation with 100\% of the dataset.
}\label{fig:100-percent-cross-project}
\begin{tabular}{lrrr}
\toprule
Model    & Mixed-project F1 & Cross-project F1 & $\Delta$ F1 \\
\midrule
LineVul  & \textbf{84.07}            & \textbf{71.81}            & -12.26   \\
Devign   & 24.32            & 14.26            & \textbf{-10.06}   \\
ReVeal   & 28.24            & 6.83             & -21.41   \\
ReGVD    & 33.61            & 21.70            & -11.91   \\
CodeBERT & 24.14            & 4.98             & -19.16   \\ \bottomrule
                               &                                       &                                       &                               \\
                               &                                       &                                       &                               \\
                               &                                       &                                       &                               \\
                               &                                       &                                       &                               \\
                               &                                       &                                       &                               \\
                               &                                       &                                       &                              
\end{tabular}
\end{table}

\end{appendices}

\end{document}